\documentclass[modern,twocolumn,twocolappendix]{aastex631}
\pdfoutput=1

\usepackage{natbib}
\usepackage{xspace}
\usepackage{amsmath}
\usepackage{lipsum}
\usepackage{appendix}

\newcommand{\msun}{$M_{\odot}$\xspace}
\newcommand{\rsun}{$R_{\odot}$\xspace}
\newcommand{\lsun}{$\mathrm{L_{\odot}}$\xspace}

\newcommand{\mearth}{$M_{\earth}$\xspace}
\newcommand{\rearth}{$R_{\earth}$\xspace}

\newcommand{\rjup}{$R_\mathrm{Jup}$\xspace}
\newcommand{\mstar}{\ensuremath{M_{\star}}\xspace}
\newcommand{\rstar}{\ensuremath{R_{\star}}\xspace}
\newcommand{\lstar}{\ensuremath{L_{\star}}\xspace}
\newcommand{\fbol}{\ensuremath{F_{\rm{bol}}}\xspace}
\newcommand{\rhostar}{\ensuremath{\rho_{\star}}\xspace}
\newcommand{\mpl}{\ensuremath{M_\mathrm{p}}\xspace}
\newcommand{\rpl}{\ensuremath{R_\mathrm{p}}\xspace}

\newcommand{\teff}{\ensuremath{T_{\mathrm{eff}}}\xspace}
\newcommand{\teffh}{\ensuremath{T_{\mathrm{eff,H}}}\xspace}
\newcommand{\teffc}{\ensuremath{T_{\mathrm{eff,C}}}\xspace}
\newcommand{\logg}{\ensuremath{\log g}\xspace}
\newcommand{\feh}{\ensuremath{[\mbox{Fe}/\mbox{H}]}\xspace}
\newcommand{\vsini}{\ensuremath{v \sin i}\xspace}
\newcommand{\gcc}{g\,cm$^{-3}$\xspace}

\newcommand{\mps}{$\mathrm{m\,s^{-1}}$\xspace}

\newcommand{\kepler}{\textit{Kepler}\xspace}
\newcommand{\ktwo}{\textit{K2}\xspace}

\newcommand{\tess}{\textit{TESS}\xspace}
\newcommand{\jwst}{\textit{JWST}\xspace}

\newcommand{\gaia}{\textit{Gaia}\xspace}

\newcommand{\isochrones}{{\textsc{isochrones}}\xspace}
\newcommand{\vespa}{{\textsc{Vespa}}\xspace}
\newcommand{\triceratops}{{\textsc{Triceratops}}\xspace}

\newcommand{\maxrad}{{\tt maxrad}\xspace}

\newcommand{\banyan}{{\tt banyan $\Sigma$}\xspace}

\newcommand{\mrexo}{{\tt MRExo}\xspace}
\newcommand{\aij}{{\tt AstroImage}J\xspace}
\newcommand{\trilegal}{\textsc{Trilegal}}{\xspace}
\newcommand{\target}{TOI-1696\xspace}
\newcommand{\ticid}{470381900\xspace}
\newcommand{\twomassid}{J04210733+4849116\xspace}
\newcommand{\wiseid}{J042107.34+484911.5\xspace}
\newcommand{\ucacid}{695-028795\xspace}

\newcommand{\steff}{$3185 \pm 76$\xspace}
\newcommand{\slogg}{$4.959 \pm 0.026$\xspace}
\newcommand{\sfeh}{$0.336 \pm 0.060$\xspace}
\newcommand{\smass}{$0.255 \pm 0.0066$\xspace}
\newcommand{\srad}{$0.2775 \pm 0.0080$\xspace}
\newcommand{\srho}{$16.8 ^{+1.5}_{-1.4}$\xspace}
\newcommand{\sdist}{$65.03 \pm 0.36$\xspace}
\newcommand{\slumi}{$0.00711 ^{+0.00083}_{-0.00075}$\xspace}
\newcommand{\ptsm}{$105.6$\xspace}
\newcommand{\pmass}{$8$\xspace}
\newcommand{\prad}{$3.09 \pm 0.11$\xspace}
\newcommand{\pper}{$2.50031 \pm 0.00001$\xspace}
\newcommand{\rvmasslim}{$48.8$\xspace}
\newcommand{\pradvalue}{$3.09$\xspace}
\newcommand{\ppervalue}{$2.5$\xspace}

\newcommand{\abc}{Astrobiology Center, 2-21-1 Osawa, Mitaka, Tokyo 181-8588, Japan}
\newcommand{\naoj}{National Astronomical Observatory of Japan, 2-21-1 Osawa, Mitaka, Tokyo 181-8588, Japan}
\newcommand{\hongo}{Department of Astronomy,  Graduate School of Science, The University of Tokyo, 7-3-1 Hongo, Bunkyo, Tokyo 113-0033, Japan}
\newcommand{\sokendai}{Department of Astronomy, School of Science, The Graduate University for Advanced Studies (SOKENDAI), 2-21-1 Osawa, Mitaka, Tokyo, Japan}
\newcommand{\eps}{Department of Earth and Planetary Science, Graduate School of Science, The University of Tokyo, 7-3-1 Hongo, Bunkyo, Tokyo 113-0033, Japan}
\newcommand{\komaba}{Department of Multi-Disciplinary Sciences, Graduate School of Arts and Sciences, The University of Tokyo, 3-8-1
Komaba, Meguro, Tokyo 153-8902, Japan}
\newcommand{\iac}{Instituto de Astrof\'\i sica de Canarias (IAC), 38205 La Laguna, Tenerife, Spain}
\newcommand{\laguna}{Departamento de Astrof\'{i}sica, Universidad de La Laguna (ULL), 38206 La Laguna, Tenerife, Spain}

\received{March 3, 2022}
\submitjournal{ApJ}

\shorttitle{TOI-1696 Validation}
\shortauthors{Mori et al.}

\graphicspath{{./}{figures/}}

\begin{document}

\title{TOI-1696: a nearby M4 dwarf with a 3\rearth planet in the Neptunian desert}

\correspondingauthor{Mayuko Mori}
\email{mori@astron.s.u-tokyo.ac.jp}

\author[0000-0003-1368-6593]{Mayuko Mori}
\affil{\hongo}

\author[0000-0002-4881-3620]{John~H.~Livingston}
\affil{\hongo}
\affil{\abc}
\affil{\naoj}

\author[0000-0002-4881-3620]{Jerome de Leon}
\affil{\hongo}

\author[0000-0001-8511-2981]{Norio Narita}
\affil{Komaba Institute for Science, The University of Tokyo, 3-8-1 Komaba, Meguro, Tokyo 153-8902, Japan}
\affil{\abc}
\affil{\iac}

\author[0000-0003-3618-7535]{Teruyuki Hirano}
\affil{\abc}
\affil{\naoj}
\affil{\sokendai}

\author[0000-0002-4909-5763]{Akihiko Fukui}
\affil{Komaba Institute for Science, The University of Tokyo, 3-8-1 Komaba, Meguro, Tokyo 153-8902, Japan}
\affil{\iac}

\author[0000-0001-6588-9574]{Karen A.\ Collins}
\affil{Center for Astrophysics \textbar \ Harvard \& Smithsonian, 60 Garden Street, Cambridge, MA 02138, USA}

\author[0000-0002-5791-970X]{Naho Fujita}
\affil{Department of Astronomy, Kyoto University, Kitashirakawa-Oiwake-cho, Sakyo-ku, Kyoto 606-8502, Japan}

\author[0000-0003-4676-0251]{Yasunori Hori}
\affil{\abc}
\affil{\naoj}

\author[0000-0001-6309-4380]{Hiroyuki Tako Ishikawa}
\affil{\abc}
\affil{\naoj}

\author[0000-0003-1205-5108]{Kiyoe Kawauchi}
\affil{\iac}
\affil{\laguna}

\author[0000-0002-3481-9052]{Keivan G.\ Stassun}
\affil{Department of Physics and Astronomy, Vanderbilt University, Nashville, TN 37235, USA}

\author[0000-0002-7522-8195]{Noriharu Watanabe}
\affil{\komaba}

\author[0000-0002-8965-3969]{Steven Giacalone}
\affil{Department of Astronomy, University of California Berkeley, Berkeley, CA 94720, USA}

\author{Rebecca Gore}
\affil{Department of Astronomy, University of California Berkeley, Berkeley, CA 94720, USA}

\author{Ashley Schroeder}
\affil{Department of Astronomy, University of California Berkeley, Berkeley, CA 94720, USA}

\author[0000-0001-8189-0233]{Courtney~D.~Dressing}
\affil{Department of Astronomy, The University of California, Berkeley, CA 94720, USA}

\author[0000-0001-6637-5401]{Allyson Bieryla}
\affil{Center for Astrophysics \textbar \ Harvard \& Smithsonian, 60 Garden Street, Cambridge, MA 02138, USA}

\author[0000-0002-4625-7333]{Eric L.\ N.\ Jensen}
\affil{Department of Physics \& Astronomy, Swarthmore College, Swarthmore PA 19081, USA}

\author[0000-0001-8879-7138]{Bob Massey}
\affil{Villa '39 Observatory, Landers, CA 92285, USA}

\author[0000-0002-1836-3120]{Avi Shporer}
\affil{Department of Physics and Kavli Institute for Astrophysics and Space Research, Massachusetts Institute of Technology, Cambridge, MA 02139, USA}

\author[0000-0002-4677-9182]{Masayuki Kuzuhara}
\affil{\abc}
\affil{\naoj}

\author[0000-0002-9003-484X]{David~Charbonneau}
\affil{Center for Astrophysics \textbar \ Harvard \& Smithsonian, 60 Garden Street, Cambridge, MA 02138, USA}

\author[0000-0002-5741-3047]{David~ R.~Ciardi}
\affil{Caltech/IPAC-NASA Exoplanet Science Institute, 770 S. Wilson Avenue, Pasadena, CA 91106, USA}

\author{John~P.~Doty}
\affil{Noqsi Aerospace Ltd., 15 Blanchard Avenue, Billerica, MA 01821, USA}

\author[0000-0002-2341-3233]{Emma Esparza-Borges}
\affil{\iac}
\affil{\laguna}

\author{Hiroki Harakawa}
\affil{Subaru Telescope, 650 N. Aohoku Place, Hilo, HI 96720, USA}

\author[0000-0003-0786-2140]{Klaus Hodapp}
\affil{University of Hawaii, Institute for Astronomy, 640 N. Aohoku Place, Hilo, HI 96720, USA}

\author[0000-0002-5658-5971]{Masahiro Ikoma}
\affil{\naoj}
\affil{\sokendai}

\author[0000-0002-5978-057X]{Kai Ikuta}
\affil{\komaba}

\author[0000-0002-6480-3799]{Keisuke Isogai}
\affil{Okayama Observatory, Kyoto University, 3037-5 Honjo, Kamogatacho, Asakuchi, Okayama 719-0232, Japan}
\affil{\komaba}

\author[0000-0002-4715-9460]{Jon~M.~Jenkins}
\affil{NASA Ames Research Center, Moffett Field, CA 94035, USA}

\author[0000-0002-5331-6637]{Taiki Kagetani}
\affil{\komaba}

\author{Tadahiro Kimura}
\affil{\eps}

\author[0000-0001-9032-5826]{Takanori Kodama}
\affil{Komaba Institute for Science, The University of Tokyo, 3-8-1 Komaba, Meguro, Tokyo 153-8902, Japan}

\author{Takayuki Kotani}
\affil{\abc}
\affil{\naoj}
\affil{\sokendai}

\author[0000-0003-2310-9415]{Vigneshwaran Krishnamurthy}
\affil{\abc}
\affil{\naoj}

\author[0000-0002-9294-1793]{Tomoyuki Kudo}
\affil{Subaru Telescope, 650 N. Aohoku Place, Hilo, HI 96720, USA}

\author{Seiya Kurita}
\affil{\eps}

\author{Takashi Kurokawa}
\affil{\abc}
\affil{Institute of Engineering, Tokyo University of Agriculture and Technology, 2-24-16, Naka-cho, Koganei, Tokyo, 184-8588, Japan}

\author[0000-0001-9194-1268]{Nobuhiko Kusakabe}
\affil{\abc}
\affil{\naoj}

\author[0000-0001-9911-7388]{David~W.~Latham}
\affiliation{Center for Astrophysics \textbar \ Harvard \& Smithsonian, 60 Garden Street, Cambridge, MA 02138, USA}

\author{Brian~McLean}
\affil{Space Telescope Science Institute, 3700 San Martin Drive, Baltimore, MD, 21218, USA}

\author[0000-0001-9087-1245]{Felipe Murgas}
\affil{\iac}
\affil{\laguna}

\author[0000-0001-9326-8134]{Jun Nishikawa}
\affil{\naoj}
\affil{\sokendai}
\affil{\abc}

\author[0000-0003-1510-8981]{Taku Nishiumi}
\affil{\sokendai}
\affil{\abc}
\affil{\komaba}

\author{Masashi Omiya}
\affil{\abc}
\affil{\naoj}

\author[0000-0002-4047-4724]{Hugh~P.~Osborn}
\affil{Department of Physics and Kavli Institute for Astrophysics and Space Research, Massachusetts Institute of Technology, Cambridge, MA 02139, USA}
\affil{NCCR/Planet-S, Universität Bern, Gesellschaftsstrasse 6, 3012 Bern, Switzerland}

\author[0000-0003-0987-1593]{Enric Palle}
\affil{\iac}
\affil{\laguna}

\author[0000-0001-5519-1391]{Hannu Parviainen}
\affil{\iac}
\affil{\laguna}

\author[0000-0003-2058-6662]{George R. Ricker} 
\affil{Department of Physics and Kavli Institute for Astrophysics and Space Research, Massachusetts Institute of Technology, Cambridge, MA 02139, USA}

\author[0000-0002-6892-6948]{Sara Seager}
\affil{Department of Earth, Atmospheric, and Planetary Sciences, Massachusetts Institute of Technology, Cambridge, MA 02139, USA}
\affil{Department of Physics and Kavli Institute for Astrophysics and Space Research, Massachusetts Institute of Technology, Cambridge, MA 02139, USA}
\affil{Department of Aeronautics and Astronautics, Massachusetts Institute of Technology, Cambridge, MA 02139, USA}

\author{Takuma Serizawa}
\affil{Institute of Engineering, Tokyo University of Agriculture and Technology, 2-24-16, Naka-cho, Koganei, Tokyo, 184-8588, Japan}
\affil{\naoj}

\author[0000-0002-6510-0681]{Motohide Tamura}
\affil{\hongo}
\affil{\abc}
\affil{\naoj}

\author[0000-0003-3860-6297]{Huan-Yu Teng}
\affil{Department of Earth and Planetary Sciences, School of Science, Tokyo Institute of Technology, 2-12-1 Ookayama, Meguro-ku, Tokyo 152-8551, Japan}

\author[0000-0003-2887-6381]{Yuka Terada}
\affil{Institute of Astronomy and Astrophysics, Academia Sinica, P.O. Box 23-141, Taipei 10617, Taiwan, R.O.C.}
\affil{Department of Astrophysics, National Taiwan University, Taipei 10617, Taiwan, R.O.C.}

\author[0000-0002-6778-7552]{Joseph~D.~Twicken}
\affil{NASA Ames Research Center, Moffett Field, CA 94035, USA}
\affil{SETI Institute, Mountain View, CA 94043, USA}

\author{Akitoshi Ueda}
\affil{\abc}
\affil{\naoj}
\affil{\sokendai}

\author[0000-0001-6763-6562]{Roland~Vanderspek}
\affiliation{Department of Physics and Kavli Institute for Astrophysics and Space Research, Massachusetts Institute of Technology, Cambridge, MA 02139, USA}

\author{Sébastien Vievard}
\affil{Subaru Telescope, 650 N. Aohoku Place, Hilo, HI 96720, USA}
\affil{\abc}

\author[0000-0002-4265-047X]{Joshua N. Winn}
\affil{Department of Astrophysical Sciences, Princeton University, 4 Ivy Lane, Princeton, NJ 08544, USA}

\author[0000-0002-5609-4427]{Yujie Zou}
\affil{\komaba}

\begin{abstract}

We present the discovery and validation of a temperate sub-Neptune around the nearby mid-M dwarf TIC 470381900 (\target), with a radius of \prad~\rearth and an orbital period of \ppervalue days, using a combination of \tess and follow-up observations using ground-based telescopes. Joint analysis of multi-band photometry from \tess, MuSCAT, MuSCAT3, Sinistro, and KeplerCam confirmed the transit signal to be achromatic as well as refined the orbital ephemeris. High-resolution imaging with Gemini/'Alopeke and high-resolution spectroscopy with the Subaru/IRD confirmed that there are no stellar companions or background sources to the star. The spectroscopic observations with IRD and IRTF/SpeX were used to determine the stellar parameters, and found the host star is an M4 dwarf with an effective temperature of \teff\,$=$\,\steff\,K and a metallicity of \feh\,$=$\, \sfeh\,dex. The radial velocities measured from IRD set a 2-$\sigma$ upper limit on the planetary mass to be \rvmasslim $M_\oplus$.
The large radius ratio (\rpl/\rstar\,$\sim$\,0.1) and the relatively bright NIR magnitude ($J$=12.2 mag) make this planet an attractive target for further followup observations. \target\,b is one of the planets belonging to the Neptunian desert with the highest transmission spectroscopy metric discovered to date, making it an interesting candidate for atmospheric characterizations with \jwst. 
\end{abstract}


\section{Introduction} \label{sec:intro}

Exoplanet population statistics from the \kepler mission \citep{Borucki2010} revealed that there is a dearth of planets around the size of Neptune ($\sim$3–4\rearth) with orbital periods less than 2–4 d. This has been referred to as the ``Neptunian Desert'' or ``photo-evaporation desert'' or simply ``evaporation desert'' \citep{Szabo2011,Mazeh2016,2017MNRAS.472..245L}. 
The scarcity of planets in this region of parameter space can be explained by photo-evaporation, that is, atmospheric mass loss due to high-energy irradiation from the host star \citep{Owen2017}. The small number of planets that have so far been found in the desert \citep[e.g.][]{West2019,Jenkins2020} are believed to retain substantial atmospheres (or are still in the process of losing them), but the physical mechanisms are not well understood. Comparing planets that have lost their atmospheres with those that have retained their atmospheres will be useful to understand the processes such as photo-evaporation theory. Therefore, it is important to increase the number of planets in this region and reveal the nature of their atmospheres. \tess \citep{Ricker2015}, which has identified over 5000 exoplanet candidates so far\footnote{As of 2022 February per \url{https://exoplanetarchive.ipac.caltech.edu/}}, made it possible to discover more planets in the Neptunian Desert.

In this paper, we report the validation of a new planet around the mid-M dwarf \target, whose transits were identified by the \tess mission. The planet  \target\,b has a sub-Neptune size (\prad~\rearth) and an orbital period of \ppervalue days, which places it within (or near the boundaries of) the Neptunian desert.

The large radius ratio (\rpl/\rstar\,$\sim$\,0.1) makes the planet's transits deep, and combined with the relatively bright near-IR (NIR) magnitude (J=12.2 mag) of the star, the planet is one of the best targets for future atmospheric research via transmission spectroscopy.

The rest of this paper is organized as follows. In Section~\ref{sec:data}, we present the observational data and the reduction procedures used for the analyses. In Section~\ref{sec:analyses}, we explain the analyses methods and results. In Section~\ref{sec:discussion}, we discuss the features of the planet and its future observational prospects, concluding with a summary in Section~\ref{sec:conclusion}.

\section{Observations \& Data Reduction}
\label{sec:data}

\subsection{Transit photometry - \tess}

\begin{figure}
    \centering
    \includegraphics[width=0.47\textwidth]{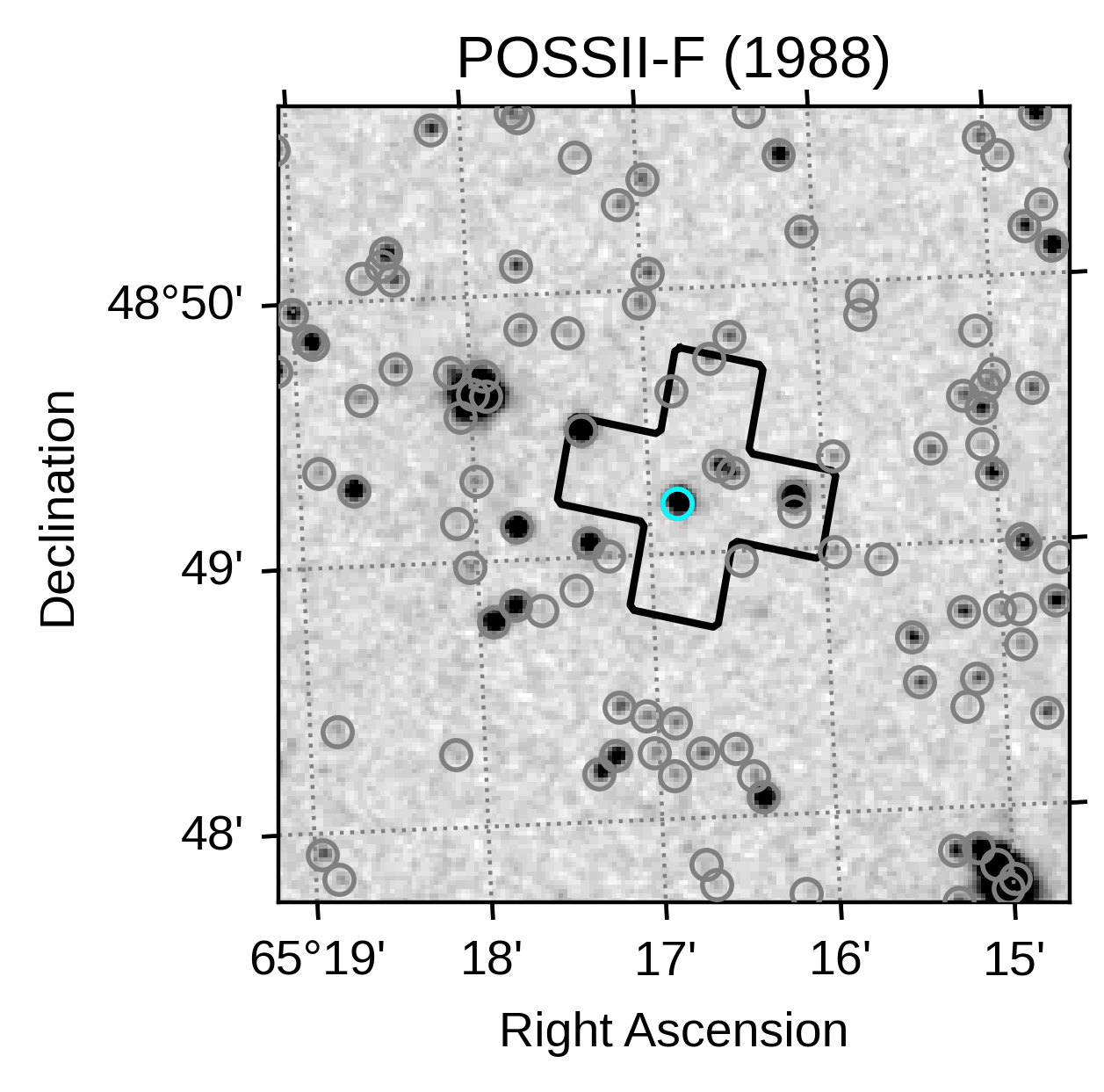}
    \caption{Archival imaging from POSSII-F survey \citep[taken in 1998;][]{1991PASP..103..661R} with the \tess photometric aperture (black outline) and \gaia sources (gray circles). The cyan circle indicates the position of \target; we note the proper motion is low enough that its current position is not significantly offset in the archival image.}
    \label{fig:aper}
\end{figure}

\begin{figure*}
    \centering
    \includegraphics[clip,trim={0 0 -1.5cm 0},width=\textwidth]{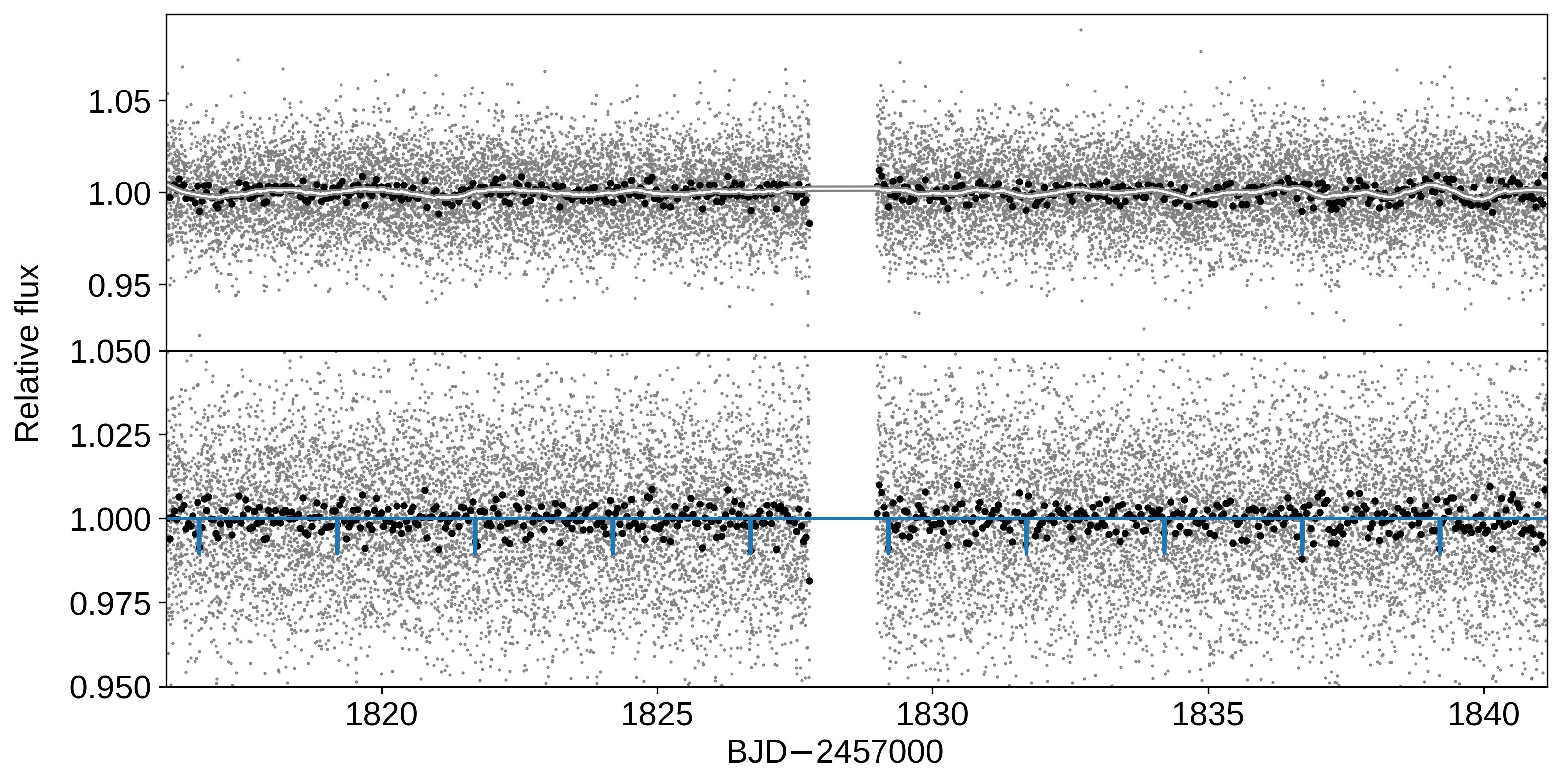}
    \includegraphics[width=\textwidth]{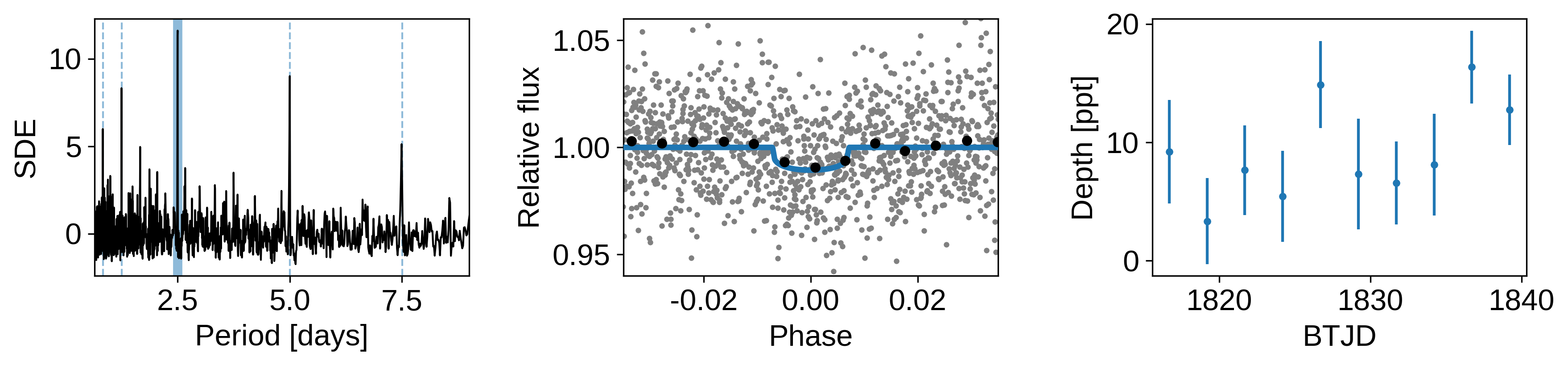}
    \caption{The upper panels show the \tess PDCSAP lightcurve with Savitzky-Golay (window=1001) variability model (top), and the flattened lightcurve with TLS model (bottom). The lower panels show the TLS power spectrum (left), folded \tess lighturve with TLS model (middle), and individual transit depths from TLS (right).}
    \label{fig:tls}
\end{figure*}

\tess observed \target with a 2 min cadence in Sector 19 from 2019 Jul 25 to Aug 22, resulting in photometry spanning approximately 27 days with a gap of about one day in the middle when the satellite reoriented itself for data downlink near perigee. Light curves were produced by the Science Processing Operations Center (SPOC) photometry pipeline \citep{Jenkins2002,Jenkins2010,Jenkins2020SPOC} using the aperture shown in Figure~\ref{fig:aper}. We used the {\tt PDCSAP} light curves produced by the SPOC pipeline \citep{2012PASP..124..985S,2012PASP..124.1000S,2014PASP..126..100S} for our transit analyses. \target is located in a fairly crowded field, owing to its low galactic latitude ($b=-0.81^{\circ}$). The SPOC pipeline applies a photometric dilution correction based on the {\tt CROWDSAP} metric, which we independently confirmed by computing dilution values based on Gaia DR2 magnitudes\footnote{approximating \gaia $R_\mathrm{p}$ as the \tess bandpass, and assuming a full width at half maximum (FWHM) of 25\,\arcsec}.

\target.01 was detected by the SPOC pipeline in a transiting planet search, and the candidate was subsequently reported to the community by the \tess Science Office (TSO) on 2020 January 30 via the \tess Object of Interest (TOI; \citet{2021GuerreroTOI}) Releases portal\footnote{\url{https://tess.mit.edu/toi-releases/}}. The candidate passed all data validation diagnostic tests \citep{Twicken2018} performed by the SPOC\footnote{Full vetting report available for download at \url{https://exo.mast.stsci.edu/exomast_planet.html?planet=TOI169601}}. The SPOC pipeline removed the transit signals of \target.01 from the light curve and performed a search for additional planet candidates \citep{Li2019}, but none were reported.

We independently confirmed the transit signal found by the SPOC. After removing stellar variability and residual instrumental systematics from the \texttt{PDCSAP} light curve using a 2$\mathrm{nd}$ order polynomial Savitzy-Golay filter, we searched for periodic transit-like signals using the transit least-squares algorithm \citep[TLS;][]{2019A&A...623A..39H}\footnote{\url{https://transitleastsquares.readthedocs.io/en/latest/index.html}}, resulting in the detection of \target.01 with a signal detection efficiency (SDE) of 11.6, a transit signal-to-noise ratio (SNR) of 7.4, orbital period of \pper~days, and transit depth of 10.6 parts per thousand (ppt), which is consistent with the values reported by the \tess team on ExoFOP-TESS\footnote{\url{https://exofop.ipac.caltech.edu/tess/}}. We subtracted this signal and repeated the transit search, but no additional signals with SDE above 10 were found. TLS also reports the approximate depths of each individual transit; we note that these transit depths and uncertainties are useful for diagnostic purposes only, as they are simplistically determined from the mean and standard deviation of the in-transit flux. The depths of the odd transits are within 1.5$\sigma$ of the even transits, suggesting a low probability of either signal being caused by an eclipsing binary at twice the detected period. The TLS detection is shown in Figure~\ref{fig:tls}.

\subsection{Transit photometry - FLWO/KeplerCam}

We used KeplerCam, mounted on the 1.2m telescope located at the Fred Lawrence Whipple Observatory (FLWO) atop Mt. Hopkins, Arizona, to observe a full transit on 2020 February 17. KeplerCam has a $23'.1 \times 23'.1$  field-of-view and operates in binned by 2 mode producing a pixel scale of 0.672\arcsec. Images were obtained in the $i$-band with an exposure time of 300 seconds. A total of 29 images were collected over 144 minutes. The data were reduced using standard IDL routines and photometry was performed using the \aij software package \citep{collins2017AIJ}.

\subsection{Transit photometry - LCO/SINISTRO}

We observed a full transit on 2020 November 13, using Sinistro, an optical camera mounted on a 1m telescope located at McDonald Observatory in Texas, operated by Las Cumbres Observatory \citep{LCOGT2013}. Sinistro has a $26'.5 \times 26'.5$ field of view with a pixel scale of 0.389\arcsec. We observed 62 images in total during 339 minutes, using a $V$-band filter, with an exposure time of 5 min. The data were reduced by the standard LCOGT \texttt{BANZAI} pipeline \citep{McCully2018}, and photometry was performed using \aij software.

\subsection{Transit photometry - LCO/MuSCAT3}

MuSCAT3 is a multi-band simultaneous camera installed on the 2m Faulkes Telescope North at Las Cumbres Observatory (LCO) on Haleakala, Maui \citep{Narita2020}. It has four channels, enabling simultaneous photometry in the $g$ (400–550 nm), $r$ (550–700 nm), $i$ (700–820 nm) and $z_s$ (820–920 nm) bands. Each channel has a 2048$\times$2048 pixel CCD camera with a pixel scale of 0.27\arcsec, providing a $9'.1 \times 9'.1$ field of view. We observed a full transit of TOI-1696.01 on 2020 December 23, from BJD 2459206.703523 to 2459206.827246. We took 36, 41, 89, and 131 exposures with exposure times of 300, 265, 120, and 80~s in the $g$, $r$, $i$, and $z_s$ bands, respectively.

The data reduction was conducted by the standard LCOGT \texttt{BANZAI} pipeline. Then the differential photometry was conducted by a customized aperture-photometry pipeline for MuSCAT series \citep{Fukui2011}. The optimized aperture radii are 8, 6, 10, and 8 pixels (2.16\arcsec, 1.62\arcsec, 2.7\arcsec, and 2.16\arcsec) for the $g$, $r$, $i$, and $z_s$ bands, respectively. We optimized a set of comparison stars for each band to minimize the dispersion of the light curves. For computational efficiency, and to achieve a more uniform signal-to-noise ratio (SNR), we subsequently binned the $g$, $r$, $i$, and $z_s$ data to 300, 240, 180, and 120~s, respectively.

\subsection{Transit photometry - NAOJ 188cm/MuSCAT}

We also observed a full transit with MuSCAT \citep{Narita2015}, which is installed on the 188cm telescope of National Astronomical Observatory of Japan (NAOJ) in Okayama, Japan. MuSCAT has a similar optical design as MuSCAT3 but has three CCD cameras for the $g$, $r$ and $z_s$ bands. On the night of 2021 July 28 we observed \target from BJD 2459424.228358 to 2459424.30679. At that point, the $r$-band camera was not available due to an instrumental issue, so we observed with only the $g$ and $z_s$ bands, using an exposure time of 60~s for both bands.

The data reduction and differential photometry was performed using the pipeline described in \citet{Fukui2011}. The optimized aperture radii were 4 and 6 pixels (1.44\arcsec and 2.16\arcsec) for the $g$ and $z_s$ bands, respectively. Similarly to the MuSCAT3 data, we binned the $g$ and $z_s$ data to 300 and 120~s, respectively.

\subsection{Speckle imaging - Gemini/'Alopeke} 
\label{sec:speckle}

\begin{figure*}
    \centering
    \includegraphics[width=0.7\textwidth]{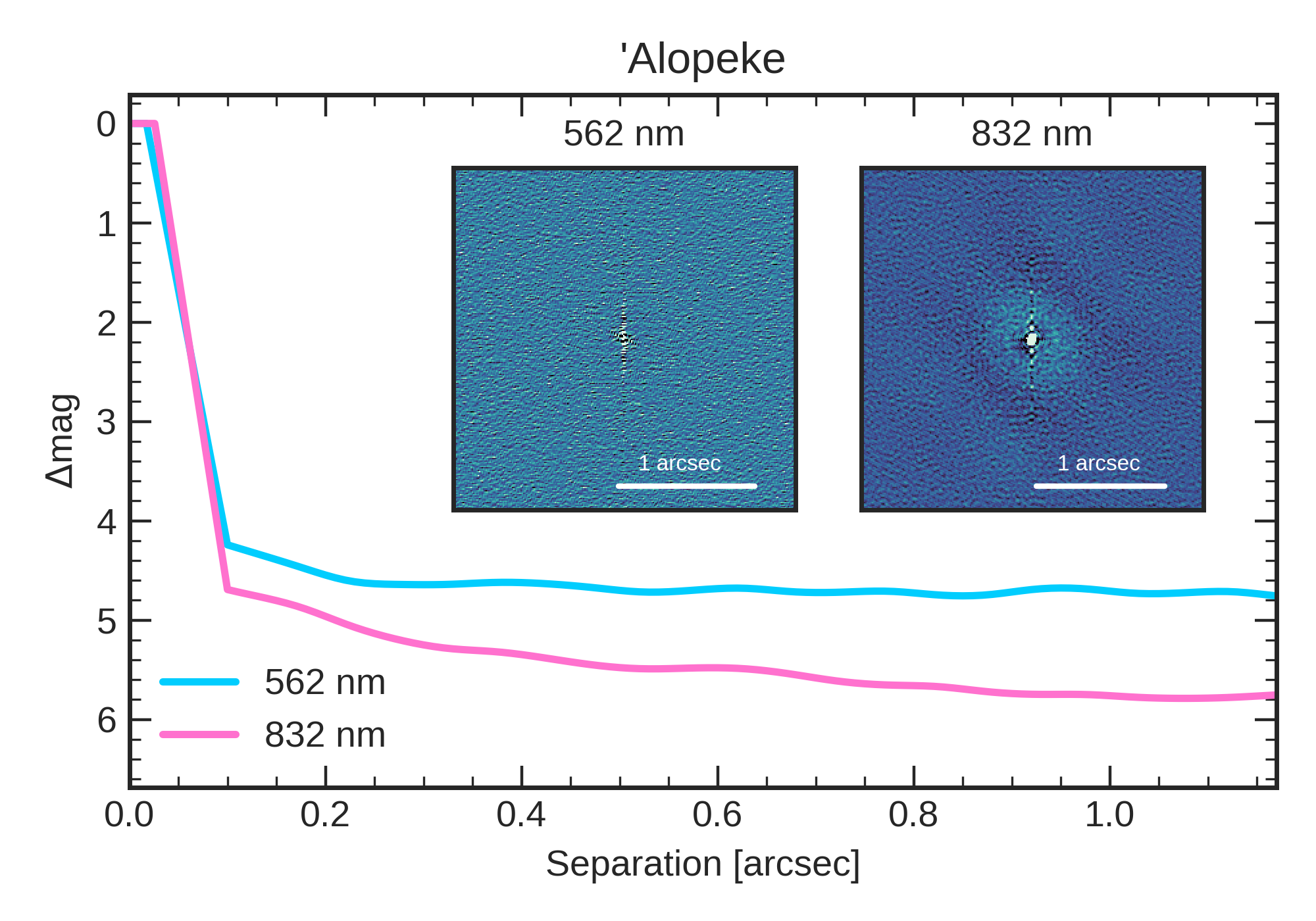}
    \caption{Gemini/'Alopeke reconstructed images and contrast curves produced as described in Section~\ref{sec:speckle}.}
    \label{fig:speckle}
\end{figure*}

On the nights of 2020 December 03 and 2021 October 14, \target was observed with the 'Alopeke speckle imager \citep{Scott2019}, mounted on the 8.1~m Gemini North telescope on Mauna Kea. 'Alopeke simultaneously acquires data in two bands centered at 562~nm and 832~nm using high speed electron-multiplying CCDs (EMCCDs). We collected and reduced the data following the procedures described in \citet{Howell2011}. The resulting reconstructed image achieved a contrast of $\Delta\mathrm{mag}=5.8$ at a separation of 1\arcsec~in the 832~nm band. No secondary sources were detected. The data taken on 2021 October 14 is shown in Figure~\ref{fig:speckle}.

\begin{figure}
    \centering
    \includegraphics[clip,trim={10cm 2cm -2cm 0},width=0.5\textwidth]{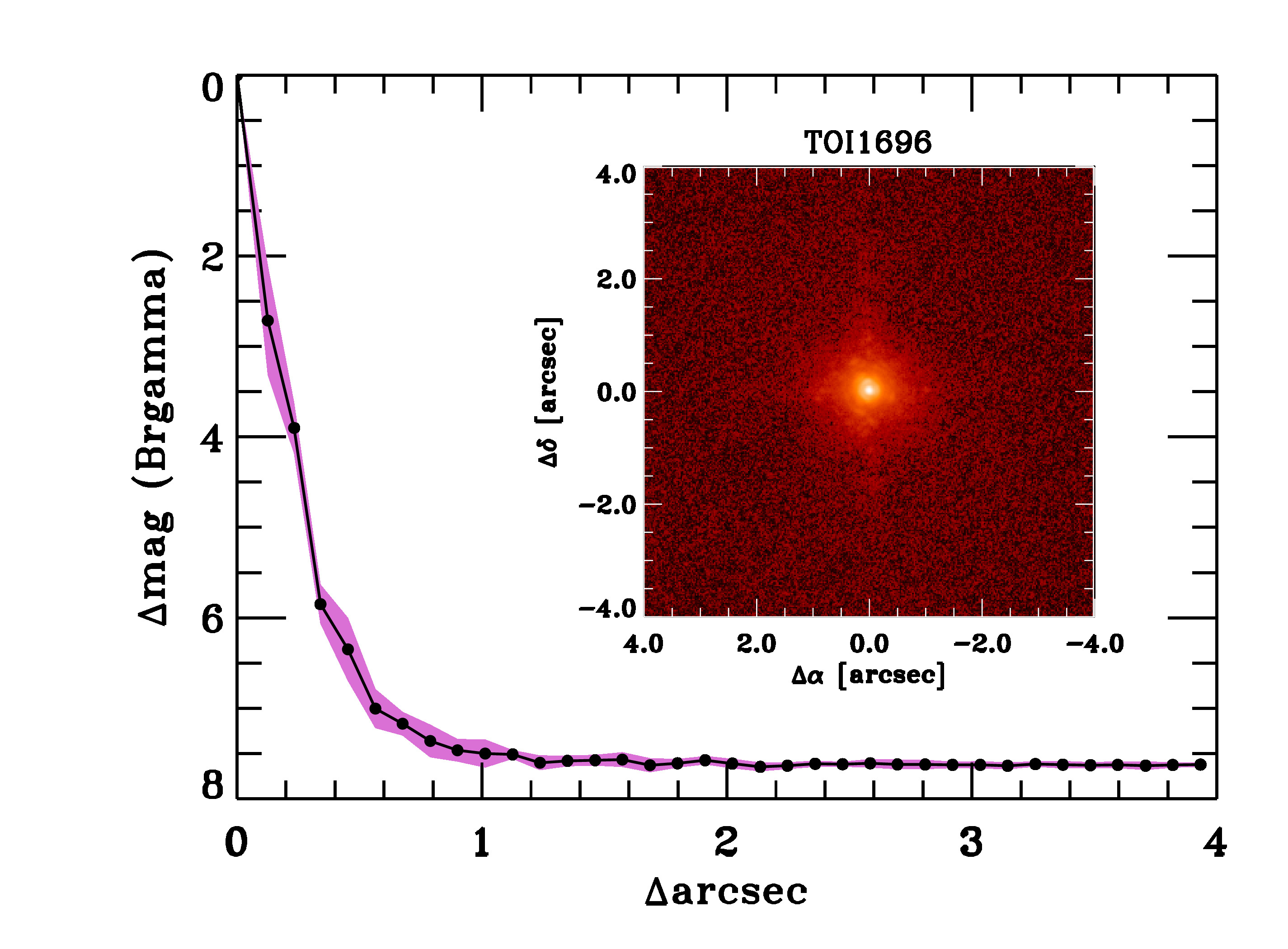}
    \includegraphics[clip,trim={10cm 2cm -2cm 0},width=0.5\textwidth]{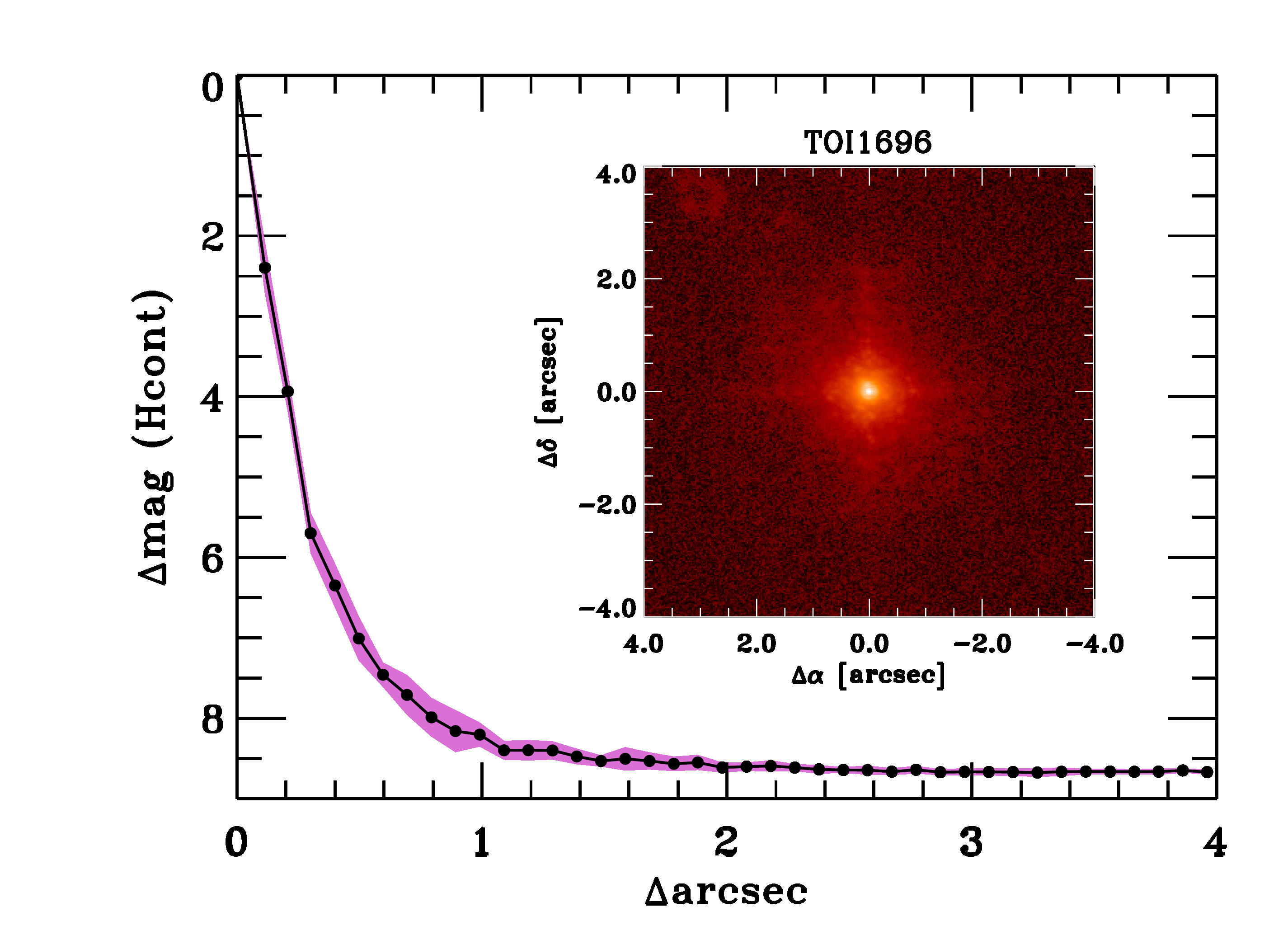}
    \caption{Palomar/PHARO images and contrast curves (top: Br$\gamma$; bottom: H$_\mathrm{cont}$) produced as described in Section~\ref{sec:ao}.}
    \label{fig:ao}
\end{figure}

\subsection{Adaptive optics imaging - Palomar/PHARO}
\label{sec:ao}

On 2021 September 19 we conducted near-infrared high-resolution imaging using the adaptive optics instrument PHARO mounted on the 5~m Hale telescope at Palomar Observatory \citep{pharo2001}. We observed \target seperately in the $Br \gamma$ ($2.18\,\rm{\mu m}$) and $H_{cont}$ ($2.29\,\rm{\mu m}$) bands, reaching a contrast of $\Delta\mathrm{mag}=8$ at a separation of 1\arcsec~in both bands. The AO images and corresponding contrast curves are shown in Figure~\ref{fig:ao}.

\subsection{High-resolution spectroscopy - Subaru/IRD}
\label{sec:obs-ird}

We obtained high-resolution spectra of \target in the NIR with IRD \citep{Tamura2012,Kotani2018}, mounted on the 8.2~m Subaru telescope. IRD can achieve a spectral resolution of $\sim$70,000 in the wavelength range 930~nm to 1740~nm. The derived spectra were used for the three purposes: to search for spectral companions (e.g. SB2 scenarios), to measure fundamental stellar parameters (e.g. effective temperature and metallicity), and to rule out large radial velocity (RV) variations that would indicate an eclipsing binary (EB), as well as placing a limit on the mass of the planet.
From UT 2021 January 30 to 2022 January 08, we obtained 13 spectra of \target using 1800~s exposure times, as part of a Subaru Intensive Program (Proposal IDs S20B-088I and S21B-118I). 
The raw data were reduced using \texttt{IRAF} \citep{Tody1993} as well as a pipeline for the detector’s bias processing and wavelength calibrations developed by the IRD instrument team \citep{Kuzuhara2018, Hirano2020PASJ}.
For the RV analyses and stellar parameter derivation, we computed a high-SNR coadded spectrum of the target following the procedures described in \citet{Hirano2020PASJ}. 

For use as a spectral template in the analysis described in Section~\ref{sec:analyses-ccf}, we also downloaded archival IRD data of GJ 699 (Barnard's Star)\footnote{Using the Subaru-Mitaka-Okayama-Kiso-Archive (SMOKA)}, which was obtained on 2019 March 23 (HST). We reduced and calibrated the GJ 699 data following the same procedures as the \target data.

\subsection{Medium-resolution spectroscopy - IRTF/SpeX}

We collected observations of \target on UT 2020 December 09 using SpeX, a medium‐resolution spectrograph on the NASA Infrared Telescope Facility (IRTF) on Maunakea \citep{Rayner2003}. We obtained our observations in SXD mode with a $0\farcs3 \times 15$\arcsec slit, providing a spectral resolution of $R \approx 2000$ over a wavelength range 700 nm to 2550 nm. In order to remove sky background and reduce systematics, the spectra were collected using an ABBA nod pattern (with a separation of $7\farcs5$ between the A and B positions) and with the slit synced to the parallactic angle. We reduced our spectra using the \texttt{Spextool} reduction pipeline \citep{Cushing2004} and removed telluric contamination using \texttt{xtellcor} \citep{Vacca2003}. The derived spectra were used to calculate the stellar metallicity.

\section{Analyses \& Results}
\label{sec:analyses}

\subsection{Stellar parameters estimation}

In the next subsections we estimate the fundamental stellar parameters of \target. First, the stellar effective temperature \teff and metallicity \feh are derived from two independent methods; one is from the IRD spectra and the other is from the SpeX spectra and photometric relations. Second, the stellar radius \rstar, mass \mstar, and other related parameters are derived using empirical relations and the above \teff and \feh values.

\subsubsection{Estimation of \teff and \feh: from IRD spectra} 
\label{sec:ird}

We derived the effective temperature \teff and abundances of individual elements [X/H] from the coadded IRD spectrum. To avoid amplifying noise in the spectrum, we decided not to deconvolve the instrumental profile prior to these analyses.

We determined the parameters by the equivalent width comparison of individual absorption lines between the synthetic spectra and the observed ones.
For \teff estimation, 47 FeH molecular lines in the Wing-Ford band at $990-1020$ nm was used as same as in \citet{Tako2022}. We also derived the abundance of eight metal elements as described in Section~\ref{sec-ap:ird_abundances}.

We iterated the \teff estimation and the abundance analysis alternately until \teff and metallicity were consistent with each other.
First, we derived a provisional \teff assuming solar metallicity ($\mathrm{[Fe/H]} = 0$), and then we determined the individual abundances of the eight elements [X/H] using this provisional \teff. 
Second, we redetermined \teff adopting the iron abundance [Fe/H] as the input metallicity, and then we redetermined the abundances using the new \teff. 
We iterated the estimation of \teff and [Fe/H] until the final results and the results of the previous step agreed within the error margin. As a result, we derived \teff $=3156 \pm 119 ~\rm{K}$ and \feh $=0.333 \pm 0.088~\rm{dex}$.

\subsubsection{Estimation of \teff and \feh: from SpeX spectra and photometric relations}
\label{sec:SpeX}

Before analyzing our SpeX spectra, we corrected the data to the lab reference frame using \texttt{tellrv}\footnote{\url{https://github.com/ernewton/tellrv}} \citep{Newton2014, Newton2022}. We then determined metallicity with \texttt{metal}\footnote{\url{https://github.com/awmann/metal}} \citep{Mann2013}, using only the $K$-band part of the spectrum, which is historically the most reliable, although the metallicities from $H$- and $J$-band are broadly consistent.

We calculated the stellar parameters using a series of photometric relations, following the Section 4.3 of \citep{Dressing2019}. First, we calculated the luminosity of the star using the Gaia EDR3 distance \citep{StassunTorres:2021}, 2MASS $J$ magnitude, $r$ magnitude \citep[from the Carlsberg Meridian Catalogue;][]{Muinos2014}, and the metallicity-dependent $r$-$J$ bolometric correction in Table 3 of \cite{Mann2015}. Next, we calculated the radius of the star using the relation between \rstar, absolute $K$ magnitude, and \feh defined in Table 1 of \cite{Mann2015}. Lastly, we calculated \teff using the Stefan-Boltzmann law. As a result, we derived \teff  $=3207 \pm 99 ~\rm{K}$ and \feh $=0.338 \pm 0.083~\rm{dex}$.

The strong agreement in \teff and \feh between the two methods suggests a high degree of reliability of the measurements. For the following analyses, we used the weighted mean of the two respective measurements for \teff and \feh, specifically, \teff $=$ \steff~K and \feh $=$\sfeh~dex.

\subsubsection{Estimation of stellar radius and mass}
\label{sec:stellarparam-empirical}

We estimated other stellar parameters such as stellar mass \mstar, radius \rstar, surface gravity \logg, mean density \rhostar, and luminosity \lstar following the procedure described in \citet{Hirano2021}. In short, the distributions of the stellar parameters are derived from a Monte Carlo approach using a combination of several empirical relations as well as the observed and literature values.

The \rstar value was calculated through the empirical relation from \citet{Mann2015}, and \mstar from \citet{Mann2019}. In deriving the stellar parameters by Monte Carlo simulations, we adopted Gaussian distributions for \teff and \feh based on our spectroscopic analyses (see Sections~\ref{sec:ird} and \ref{sec:SpeX}), the apparent $K_s$-band magnitude from 2MASS, and the parallax from \gaia EDR3 \citep{StassunTorres:2021}. We assumed zero extinction ($A_V = 0$), considering the proximity of the star to Earth.

As a result, we derived \rstar~$=$~\srad~\rsun and \mstar~$=$~\smass~\msun along with the other parameters listed in \ref{tab:ap-stellar-comp}. By interpolating Table 5 of \citet{mamajek} we determined the spectral type of \target to be M4V (M3.9V $\pm$ 0.2). 

To check the robustness of this analysis, we confirmed them to be in good agreement with stellar parameters derived through independent analyses based on SED fitting and \isochrones (see Section~\ref{sec-ap:stellarparam-SED} and \ref{sec-ap:stellarparam-isochrones}).

\begin{table}
\scriptsize
\centering
\caption{Main identifiers, equatorial coordinates, proper motion, parallax, optical and infrared magnitudes, and fundamental parameters of \target.}
\label{tab:stellar}
\begin{tabular}{lrr}
\hline
Parameter & Value & Source \\
\hline
\multicolumn{3}{l}{\it Main identifiers}  \\
\noalign{\smallskip}
\multicolumn{2}{l}{TIC}{\ticid}  & TIC v8$^a$ \\
\multicolumn{2}{l}{2MASS}{\twomassid}  & ExoFOP$^a$ \\
\multicolumn{2}{l}{WISE}{\wiseid}  & ExoFOP$^a$ \\
\multicolumn{2}{l}{UCAC4}{\ucacid}  & ExoFOP$^a$ \\
\multicolumn{2}{l}{\gaia EDR3}{270260649602149760}  & \gaia EDR3$^b$ \\
\hline
\multicolumn{3}{l}{\it Equatorial coordinates, parallax, and proper motion}  \\
\noalign{\smallskip}
R.A. (J2015.5)	& 04$^\mathrm{h}$21$^\mathrm{m}$07.36$^\mathrm{s}$ & \gaia EDR3$^b$ \\ 
Dec. (J2015.5)	& $+$48$\degr$49$\arcmin$11.38$\arcsec$	& \gaia EDR3$^b$ \\
$\pi$ (mas) 	& $15.4752 \pm 0.0345$ & \gaia EDR3$^b$ \\
$\mu_\alpha$ (mas\,yr$^{-1}$) 	& $12.8726 \pm 0.0345$		& \gaia EDR3$^b$ \\
$\mu_\delta$ (mas\,yr$^{-1}$) 	& $-19.0463 \pm 0.0269$		& \gaia EDR3$^b$ \\
\hline
\multicolumn{3}{l}{\it Optical and near-infrared photometry} \\
\noalign{\smallskip}
$TESS$              & $13.9664 \pm 0.00730068	$     & TIC v8$^a$     \\
\noalign{\smallskip}
$G$				 & $15.3056 \pm 0.0028$	& \gaia EDR3$^b$ \\
$B_\mathrm{p}$   & $17.0511 \pm 0.0051$   & \gaia EDR3$^b$ \\
$R_\mathrm{p}$   & $14.0457 \pm 0.0039$   & \gaia EDR3$^b$ \\
\noalign{\smallskip}
$B$              & $18.467 \pm 0.162$          & ExoFOP$^a$ \\
$V$              & $16.82 \pm 1.133$          & ExoFOP$^a$ \\
\noalign{\smallskip}
$J$ 			&  $12.233 \pm 0.023$      & 2MASS$^c$ \\
$H$				&  $11.604 \pm 0.031$      & 2MASS$^c$ \\
$Ks$			&  $11.331 \pm 0.023$      & 2MASS$^c$ \\
\noalign{\smallskip}
$W1$			&  $11.134 \pm 0.023$      & All{\it WISE}$^d$ \\
$W2$			&  $10.984 \pm 0.021$      & All{\it WISE}$^d$ \\
$W3$			&  $10.71 \pm 0.11$      & All{\it WISE}$^d$ \\
$W4$			&  $8.748 \pm $      & All{\it WISE}$^d$ \\
\hline
\multicolumn{3}{l}{\it Fundamental parameters}   \\
\teff (K) & \steff & This work \\
\logg (cgs) & \slogg & This work \\
\feh (dex) & \sfeh & This work \\
\mstar (\msun) & \smass & This work \\
\rstar (\rsun) & \srad & This work \\
\rhostar (\gcc) & \srho & This work \\
distance (pc) & \sdist & This work \\
Luminosity (\lsun) & \slumi & This work \\
\hline
\noalign{\smallskip}
\multicolumn{3}{l}{$^a$\url{https://exofop.ipac.caltech.edu/tess/}} \\
\multicolumn{3}{l}{$^b$\citet{StassunTorres:2021}} \\
\multicolumn{3}{l}{$^c$\citet{2006AJ....131.1163S}} \\
\multicolumn{3}{l}{$^d$\citet{2014yCat.2328....0C}} \\
\end{tabular}
\end{table}

\subsection{Search for spectroscopic binary stars}\label{sec:analyses-ccf}

If a stellar companion orbits the target star, the observed spectra will generally be the combination of two stellar spectra with different radial velocities. 
To see if \target is a spectroscopic binary (i.e. an SB2), we calculated the cross-correlation function (CCF) of the \target's IRD spectra with that of the well-known single-star GJ 699 (Barnard's Star). The spectrum of \target used for the analysis was obtained on UT 2021 January 30 08:53, which corresponds to the an orbital phase of 0.247 based on the \tess ephemeris. 

\begin{figure}
    \centering
    \includegraphics[width=\columnwidth]{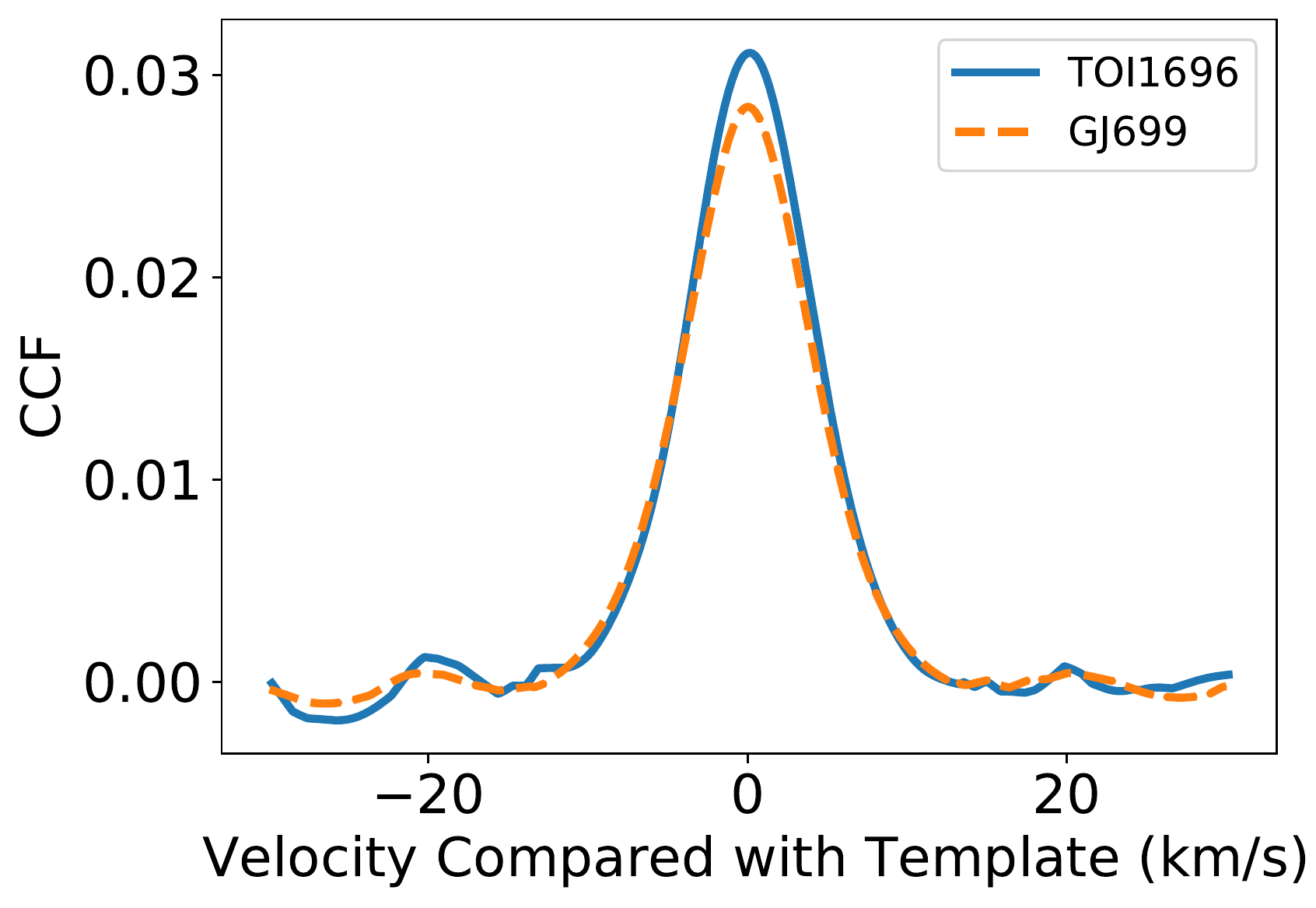}
\caption{Calculated CCF of the IRD spectrum of \target taken on 2021 January 30 at the orbital phase 0.247, to the template spectrum of GJ699, exhibiting a single peak with width 9.2~k\mps. The dashed line shows the auto-correlation function of the GJ699 spectrum as a reference.}
    \label{fig:CCF}
\end{figure}

For the analysis, we divided the spectra into six wavelength bins that are less affected by telluric absorption: [988, 993 nm], [995, 1000 nm], [1009, 1014 nm], [1016, 1021 nm], [1023, 1028 nm], and [1030, 1033 nm]. We corrected the telluric absorption signal using the spectra of the rapid-rotator HIP~74625, which was observed at the same night. The CCF to the template spectrum was calculated for each segment, after barycentric velocity correction. Finally, we computed the median of the CCFs from each segment. As shown in Figure~\ref{fig:CCF}, the resulting CCF is clearly single-peaked. If the observed transit signals were actually caused by an eclipsing stellar companion, the RV difference at quadrature would be $> 100 \,\rm{km\,s^{-1}}$, which would result in a second peak in the CCF given that the flux of such a companion would be detectable. We thus conclude \target is not an eclipsing binary.

\subsection{Stellar age}
\label{sec:analyses-age}

Because young stars are active and rapidly rotating, stellar activity and rotation period can be used as proxy for determining its youth. We did not find any stellar rotational signal in the \tess SPOC light curve, suggesting that the star is not very active. 
Similarly, no strong rotational signal was found in archival photometric data from ZTF Data Release 9 \citep{ztf1, ztf2} and ASAS-SN \citep{asassn}.

GJ 699 has a rotation period of 145 days and \vsini of less than 3 $\,\rm{km\,s^{-1}}$ \citep{Barnard}, which is below the limit of IRD's resolving power ($\sim$70000, corresponding to $\sim 4.5\,\rm{km\,s^{-1}}$). While the CCF of \target has a FWHM value consistent with that of GJ 699 (see Figure~\ref{fig:CCF}), even if we assume the rotation axis of \target is in the plane of the sky, relatively short rotation periods cannot be ruled out, as their rotational broadening would not be resolvable with IRD. However, fast rotation would most likely be accompanied with surface magnetic activity levels that would produce detectable photometric signals. 
We also used \banyan \citep{Gagne2018} to check if \target is a member of any known stellar associations, using its proper motion and the parallax from \gaia EDR3. \banyan tool\footnote{\url{http://www.exoplanetes.umontreal.ca/banyan/}} returned a value of 99.9\% field star, suggesting it is not a member of any nearby young moving group.
The non-detection by GALEX also means that the star is not young enough to be bright in the UV.
We thus conclude that \target is most likely a relatively old, slow rotator.

\subsection{Transit analysis}\label{sec:analyses-transit}

We jointly fit the \tess, KeplerCam, Sinistro, MuSCAT3, and MuSCAT datasets using the \texttt{PyMC3} \citep{pymc3}, \texttt{exoplanet}\footnote{\url{https://docs.exoplanet.codes/en/stable/}} \citep{exoplanet}, \texttt{starry} \citep{luger18}, and \texttt{celerite2} \citep{celerite1,celerite2} software packages. The model assumes a chromatic transit depth, a linear ephemeris, a circular orbit, and quadratic limb darkening. For efficient and uninformative sampling, the quadratic limb darkening coefficients were transformed following \citet{kipping13}. To account for systematics in the ground-based datasets we included a linear model of airmass and other covariates, such as the pixel response function peak, width, and centroids, when available. To account for stellar variability and residual systematics in the \tess \texttt{SPOC} light curve, we included a Gaussian Process \citep[GP][]{RasmussenWilliams2005} model with a Mat\'ern-3/2 covariance function. To account for the possibility of under- or over-estimated uncertainties, we included a white noise scale parameter for each dataset/band, enabling the errors to be estimated simultaneously with other free parameters; we placed Gaussian priors on these white noise scale parameters, with center and width equal to unity. We placed Gaussian priors on the stellar mass and radius based on the results in Table~\ref{tab:stellar}. We also placed Gaussian priors on the limb darkening coefficients based on interpolation of the parameters tabulated in \citet{Claret2012} and \citet{Claret2017}, propagating the uncertainties in the stellar parameters in Table~\ref{tab:stellar} via Monte Carlo simulation.

To optimize the model we used the gradient-based {\tt BFGS} algorithm \citep{NoceWrig06} implemented in {\tt scipy.optimize} to find initial maximum a posteriori (MAP) parameter estimates. We then used these estimates to initialize an exploration of parameter space via ``no U-turn sampling'' \citep[NUTS,][]{HoffmanGelman2014}, an efficient gradient-based Hamiltonian Monte Carlo (HMC) sampler implemented in {\tt PyMC3}. 

\begin{figure*}
    \centering
    \includegraphics[clip,trim={0 0 0 0},width=0.95\textwidth]{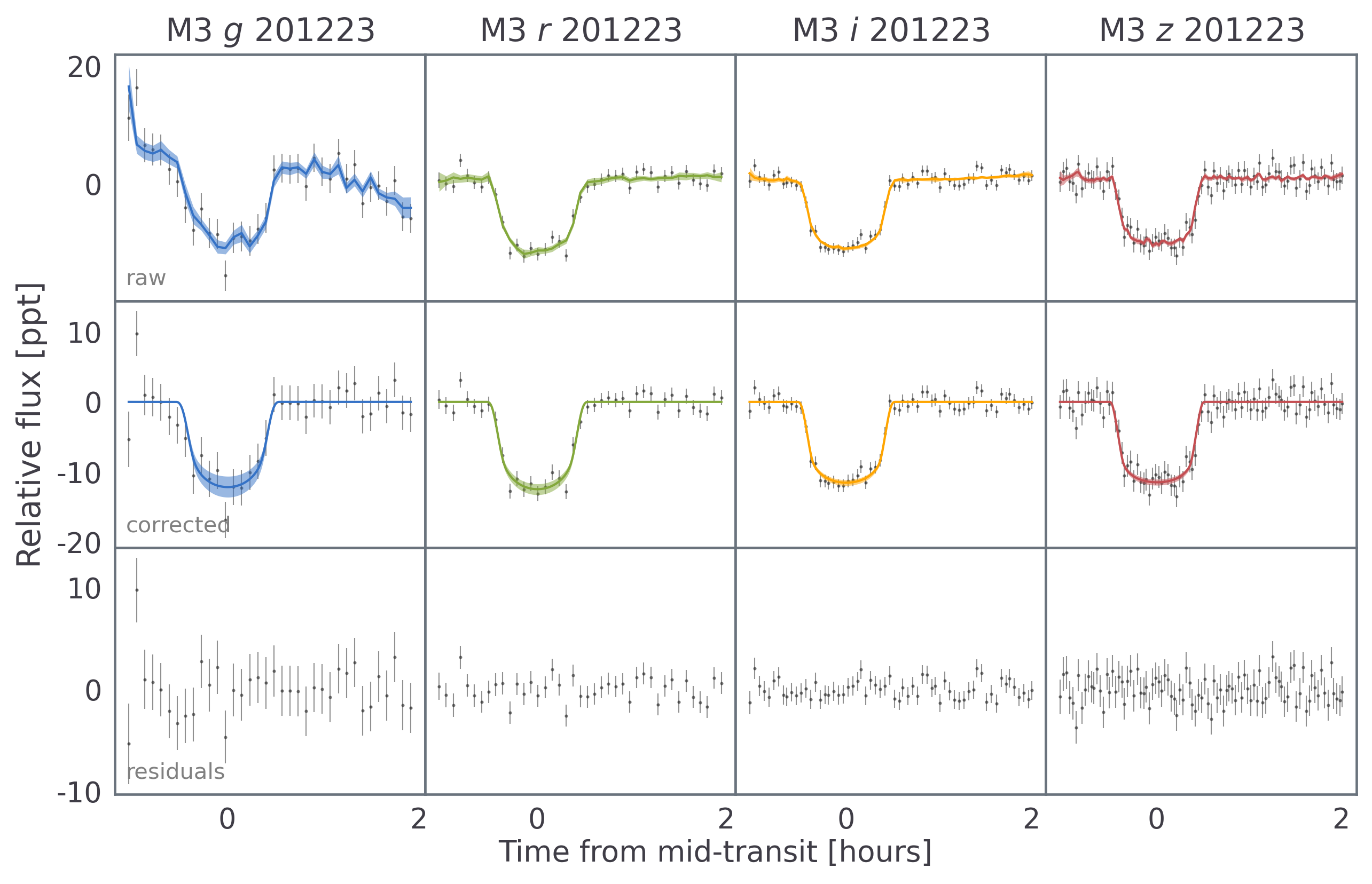}
    \caption{Transit model fit to the MuSCAT3 (M3) data from 2020 December 23, ordered column-wise per bandpass. The top row shows the raw data with the transit and systematics model, the middle row shows the systematics-corrected data with only the transit model, and the bottom row shows the residuals from the fit. The colors of the model correspond to the photometric bandpass of each dataset; see also Figure~\ref{fig:post}.}
    \label{fig:muscat3}
\end{figure*}

\begin{figure}
    \centering
    \includegraphics[clip,trim={0 0 0 0},width=0.47\textwidth]{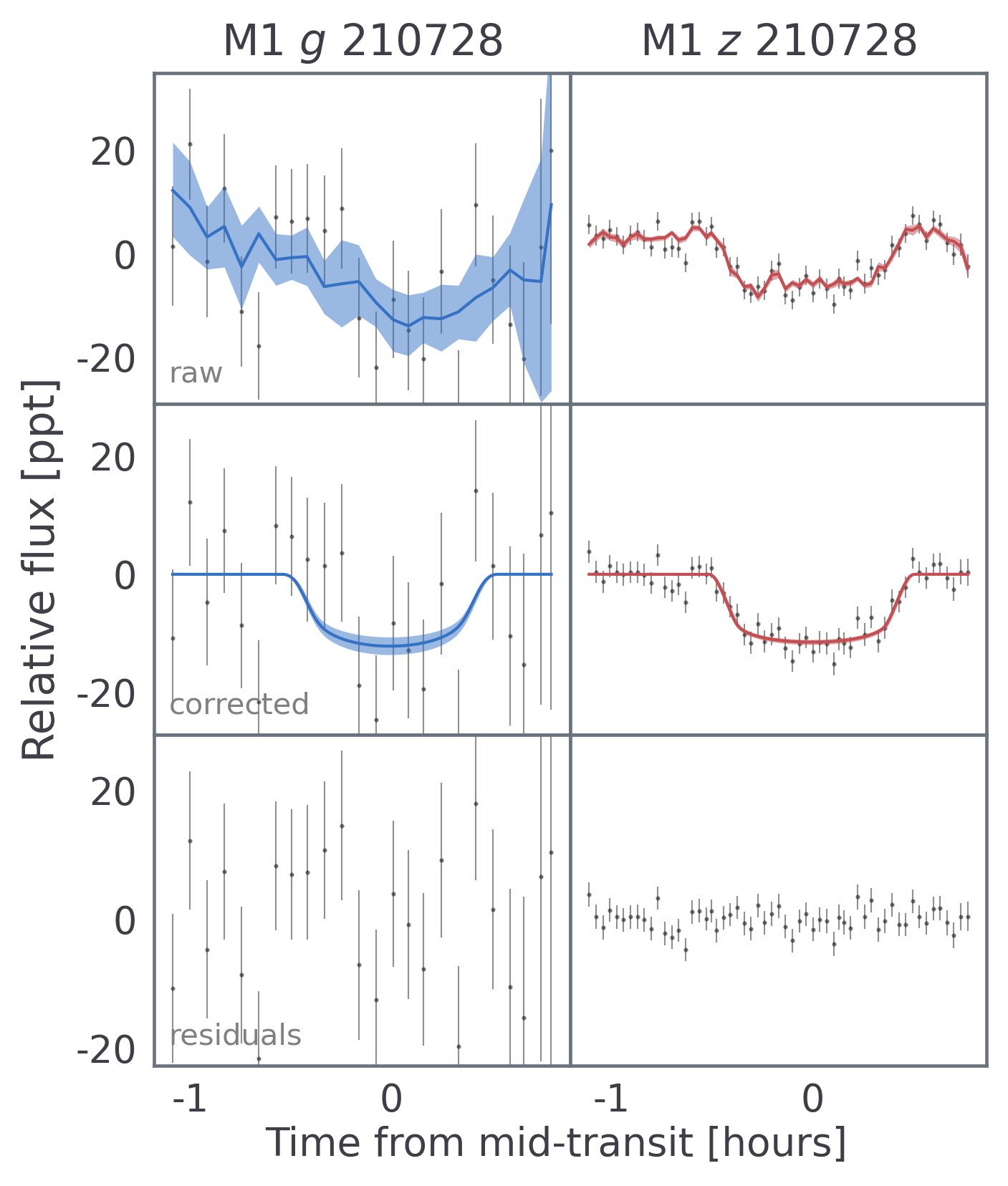}
    \caption{Same as Figure~\ref{fig:muscat3}, but for the MuSCAT (M1) data from 2021 July 28.}
    \label{fig:muscat1}
\end{figure}

\begin{figure}
    \centering
    \includegraphics[clip,trim={0 0 0 0},width=0.47\textwidth]{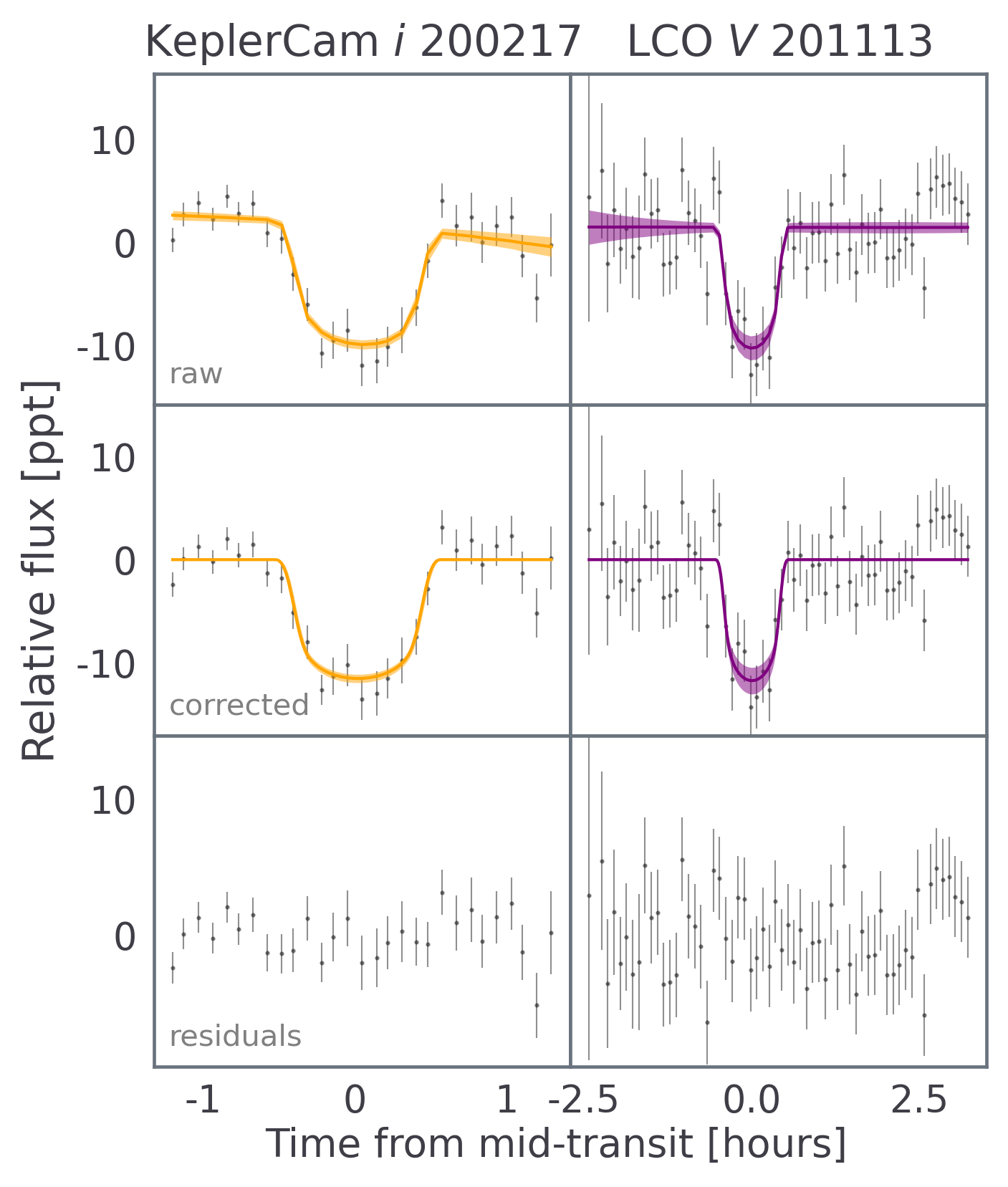}
    \caption{Same as Figure~\ref{fig:muscat3}, but for the KeplerCam and LCO data from 2020 February 17 and 2021 November 13, respectively.}
    \label{fig:kc-lco}
\end{figure}

\begin{figure*}
    \centering
    \includegraphics[width=0.7\textwidth]{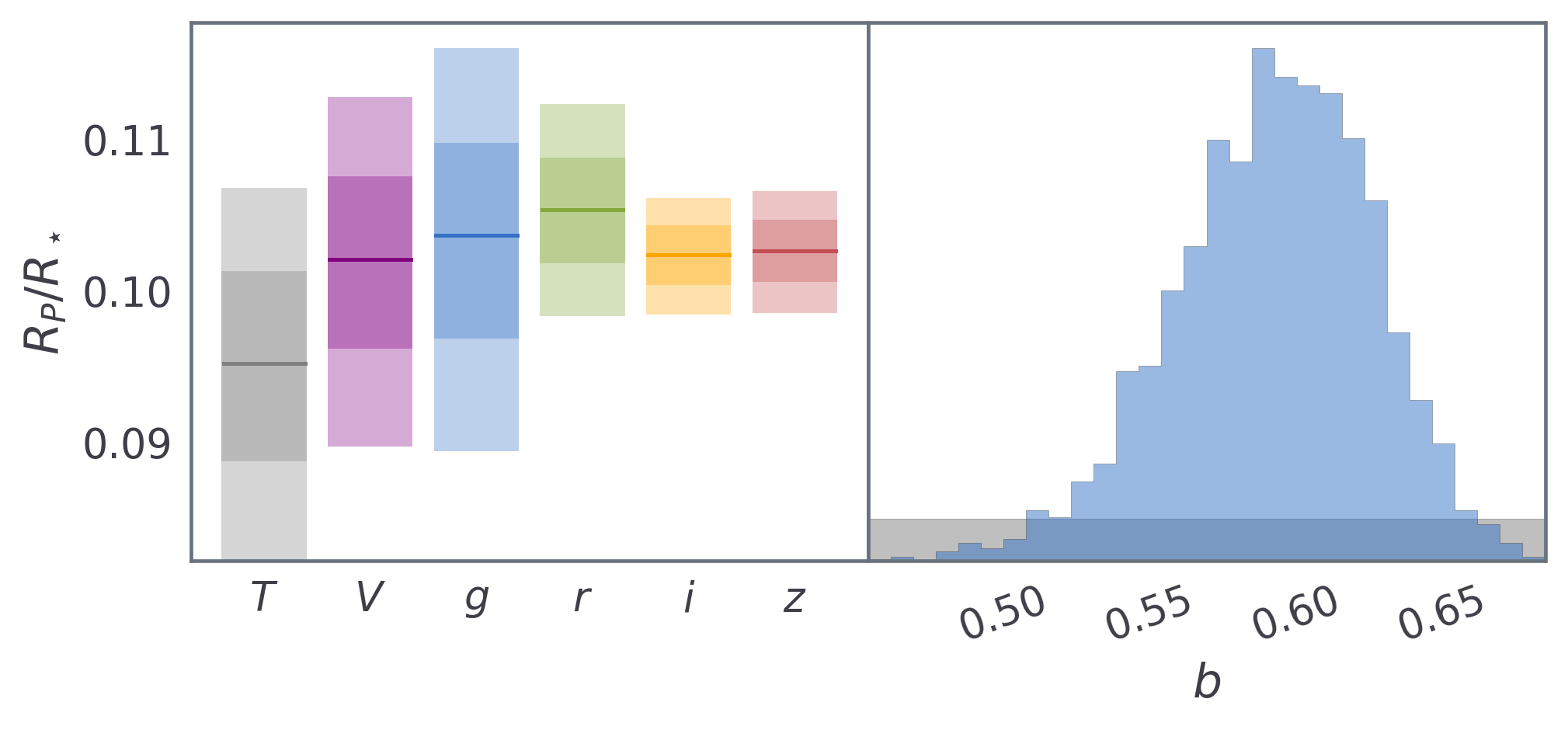}
    \caption{Posteriors of the planet-to-star radius ratio  ($R_P/R_\star$) in each bandpass (left) and impact parameter (right) from the joint fit to the \tess data and the ground-based data shown in Figure~\ref{fig:muscat3}, Figure~\ref{fig:muscat1}, and Figure~\ref{fig:kc-lco}; the gray shaded region in the right panel represents the uniform prior used in the fit, while the blue histogram is the posterior.}
    \label{fig:post}
\end{figure*}

Detailed plots showing the model fits to the ground-based datasets are shown in Figure~\ref{fig:muscat3}, Figure~\ref{fig:muscat1}, and Figure~\ref{fig:kc-lco}. We did not detect any significant wavelength dependence of the transit depth (see Figure~\ref{fig:post}), which rules out many plausible false positive scenarios involving eclipsing binaries (see Section~\ref{sec:validation} for more details). The results of this fit are listed in Table~\ref{tab:results}. Having established the achromaticity of the transit depth, we conducted a second fit with an achromatic model to robustly estimate the planet radius. This fit resulted in a final value of $R_P/R_\star = 0.1025 \pm 0.0014$, corresponding to an absolute radius of \prad~\rearth, and all other parameters were unchanged.

\begin{table}
\centering
\caption{Results of joint fit to the \tess and ground-based transit datasets. \label{tab:results}}
\begin{tabular}{lc}
Parameter & Value \\
\hline
\multicolumn{2}{l}{\it Primary transit parameters} \\
$M_\star$ [$M_\odot$] & $0.255 \pm 0.007$ \\
$R_\star$ [$R_\odot$] & $0.277 \pm 0.008$ \\
$T_0$ [BJD] & $2458834.20115 \pm 0.00058$ \\
$P$ [days] & $2.500311 \pm 0.000004$ \\
$R_P/R_{\star,T}$ & $0.0952 \pm 0.0062$ \\
$R_P/R_{\star,V}$ & $0.1021 \pm 0.0057$ \\
$R_P/R_{\star,g}$ & $0.1036^{+0.0060}_{-0.0068}$ \\
$R_P/R_{\star,r}$ & $0.1053 \pm 0.0034$ \\
$R_P/R_{\star,i}$ & $0.1023 \pm 0.0020$ \\
$R_P/R_{\star,z}$ & $0.1026 \pm 0.0020$ \\
$R_P/R_\star$ & $0.1025 \pm 0.0014^a$ \\
$b$ & $0.59^{+0.03}_{-0.04}$ \\
\hline
\multicolumn{2}{l}{\it Limb darkening parameters} \\
$u_1$ ($T$) & $0.16 \pm 0.01$ \\
$u_2$ ($T$) & $0.48 \pm 0.01$ \\
$u_1$ ($V$) & $0.48 \pm 0.02$ \\
$u_2$ ($V$) & $0.30 \pm 0.01$ \\
$u_1$ ($g$) & $0.49 \pm 0.01$ \\
$u_2$ ($g$) & $0.31 \pm 0.01$ \\
$u_1$ ($r$) & $0.50 \pm 0.01$ \\
$u_2$ ($r$) & $0.25 \pm 0.01$ \\
$u_1$ ($i$) & $0.37 \pm 0.01$ \\
$u_2$ ($i$) & $0.28 \pm 0.01$ \\
$u_1$ ($z$) & $0.24 \pm 0.01$ \\
$u_2$ ($z$) & $0.36 \pm 0.01$ \\
\hline
\multicolumn{2}{l}{\it Derived parameters} \\
$R_\mathrm{p}$ [$R_\oplus$] & $3.09 \pm 0.11^a$ \\
$a$ [AU] & $0.0229 \pm 0.0002$ \\
$T_\mathrm{eq}$ [K] & $489 \pm 13^b$ \\
$T_{14}$ [hours] & $1.00 \pm 0.01$ \\
\hline
\noalign{\smallskip}
\multicolumn{2}{l}{$^a$Derived from achromatic transit model fit.} \\
\multicolumn{2}{l}{$^b$Assuming a Bond albedo of 0.3.} \\
\end{tabular}
\end{table}

\begin{figure}
    \centering
    \includegraphics[width=0.47\textwidth]{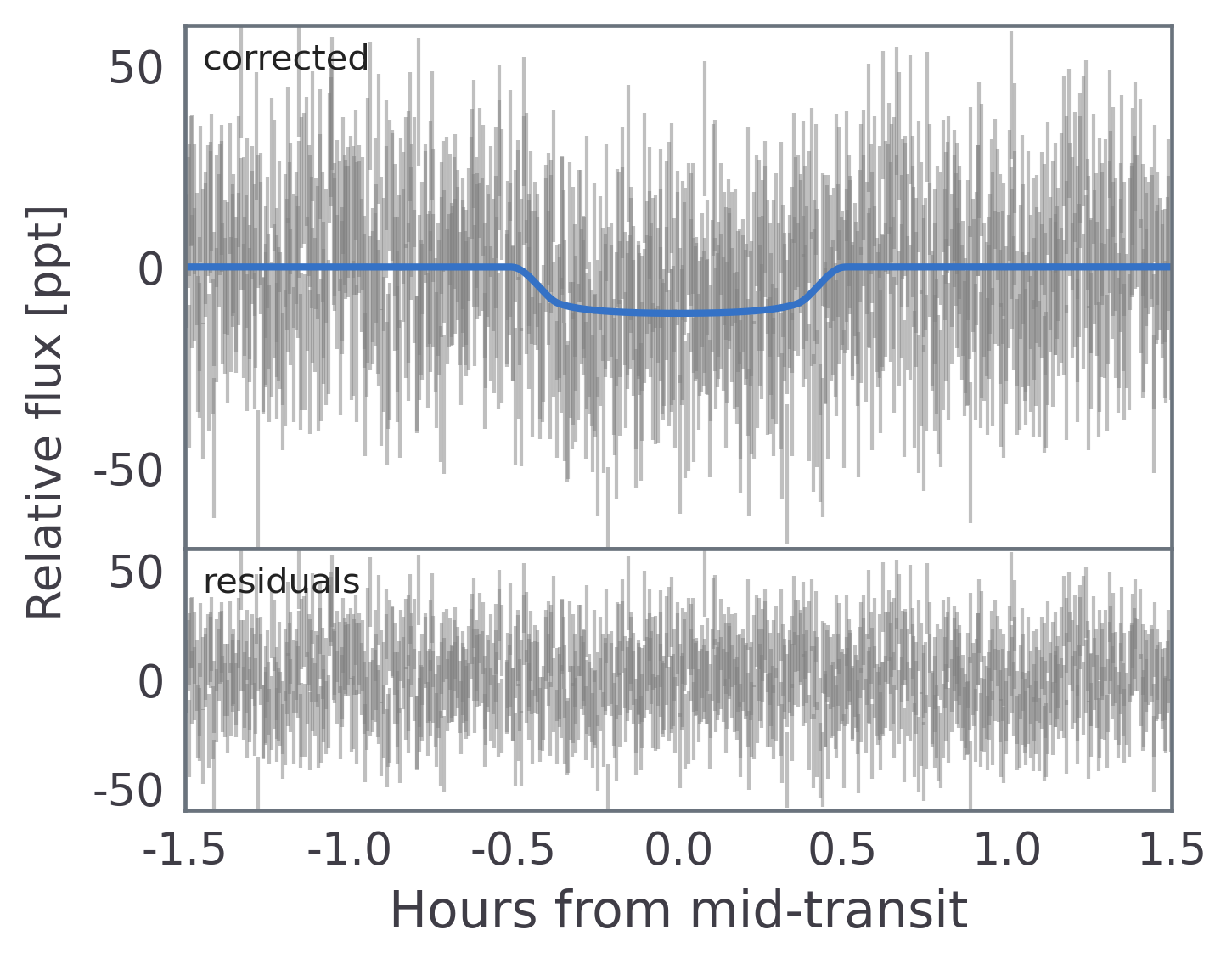}
    \caption{The phase-folded \tess light curve after removing the best-fit GP noise model, with the best-fit transit model (blue) from our joint analysis of the \tess and ground-based light curves.}
    \label{fig:tess}
\end{figure}

\subsection{Companion mass constraints}
\label{sec:mass}

To put a limit on the mass of \target.01, we fit an RV model with a circular orbit to the RV data from Subaru/IRD. Between the $H$-band and the $YJ$-band spectra obtained with IRD, we opted to use the $H$-band spectra for RV analysis because of its higher SNR.\footnote{There have been reports of unpredictable systematic errors caused by persistence light on the detector in $H$-band, especially when bright stars are observed before fainter stars. We checked the objects observed before \target and found that none were more than 1.2 mag brighter in the $H$-band, i.e. persistence light isn't likely to be a problem with these data.} The data observed on 2021 January 29 was excluded because of the possibility of an RV offset, as there was a gap of 8 months relative to the succeeding observations. We also removed any data with the clouds passing, which can cause systematic errors. The final dataset consisted of 9 RV measurements from 2021 September 29 to 2022 January 8.

We used the RV model included in PyTransit which we simplified to have five free parameters: phase-zero epoch $\rm{T_0}$, period, RV semi-amplitude, RV zero point, and RV jitter term. For the $\rm{T_0}$ and the period, we put Gaussian priors using the $\rm{T_0}$ and period derived from the transit analysis. For the other parameters we put wide uniform priors. We ran the built-in Differential Evolution optimizer and then sample the parameters with Markov Chain Monte Carlo (MCMC) using 30 walkers and 10$^4$ steps. We use the following equation to derive the planet mass,
\begin{equation}
    \centering
    M_p = \Big( \frac{PM_s^2}{2\pi } \Big)^{1/3} \frac{K(1-e^2)^{1/2}}{\sin(i)}
\end{equation}
where \mpl is planet mass, \mstar is star mass, $P$ is orbital period, $K$ is RV semi-amplitude, $e$ is eccentricity (fixed to zero), and $i$ is inclination (fixed to 90$^{\circ}$).
To propagate uncertainties, we use the posteriors for \mstar and $P$ from previous analyses. 

\begin{figure}
    \centering
    \includegraphics[width=\columnwidth]{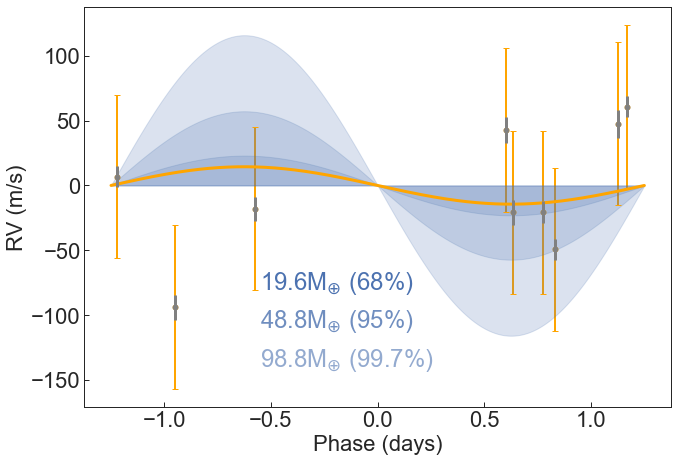}
    \caption{Phase-folded RVs with Keplerian models corresponding to the 1-, 2-, and 3-$\sigma$ mass upper limits. Gray points with the error bars show the errors estimated from the data processing method described in Section~\ref{sec:obs-ird}. The error bars in orange show the original errors + jitter term value of 62~\mps (added in quadrature) from the best-fit RV model (orange line, best $\rm{M_p}=12.3$\mearth.)
    }
    \label{fig:RV}
\end{figure}

In Figure~\ref{fig:RV} we plot Keplerian orbital models corresponding to different masses encompassing the 68$^\mathrm{th}$, 95$^\mathrm{th}$, and 99.7$^\mathrm{th}$ percentiles of the semi-amplitude posterior distribution. The 2-$\sigma$ upper limit is \rvmasslim~\mearth 
which places the companion 2 orders of magnitude below the deuterium burning mass limit. The best-fit semi-amplitude is $\rm{K}=14.4\,\rm{m s^{-1}}$, which corresponds to a mass of $\rm{M_p}=12.3$\mearth, and the best-fit jitter value is $\sigma_{K}$=62~\mps.

We calculated an expected planetary mass of $\sim$\pmass\,\mearth with \mrexo\footnote{\url{https://github.com/shbhuk/mrexo}}, which uses a mass-radius relationship calibrated for planets around M dwarfs \citep{Kanodia2019}. This mass corresponds to a semi-amplitude of 9.4\mps, but the observed RV data exhibits significantly larger variability ($\sigma\approx$~52\mps). We interpret this variability as being responsible for the large jitter value found by the fit, which suggests it is out-of-phase with \target.01. Since the star appears to be quiet, one explanation for this signal is the existence of an additional (possibly non-transiting) planet, but more RV measurements would be required to determine if this is the case. Furthermore, if such a planet were dynamically interacting with \target.01, then this could help explain \target.01's location in a sparsely populated part of the period-radius plane (see Section~\ref{sec:discussion}). 

\subsection{Eliminating false positive scenarios}
\label{sec:validation}

A number of astrophysical scenarios can mimic the transit signal detected from \tess photometry, including an eclipsing binary (EB) with a grazing transit geometry, a hierarchical EB (HEB), and a diluted eclipse of a background (or foreground) EB (BEB) along the line of sight of the target. In the following, we will examine the plausibility of each scenario. 

First, the Renormalised Unit Weight Error (RUWE) from \gaia EDR3 is 1.12, which suggests that \target is single \citep{2020Belokurov}. We can also rule out the EB scenario based on the analysis of the IRD CCF in Section~\ref{sec:analyses-ccf}, and the mass constraint derived in Section~\ref{sec:ird}.
Finally, the absence of any wavelength dependence of the transit depth from our chromatic transit analysis (Section~\ref{sec:analyses-transit}) is incompatible with contamination from a star of different spectral type (colour) than the host star, the details of which are discussed in the Appendix~\ref{app-contamination}. 
In the absence of dilution, the measured radius of \prad~\rearth (0.27~\rjup) equals the true radius, which makes it significantly smaller than the lower limit of 0.8~\rjup expected for brown dwarfs \citep{2011Burrows}. 

Grazing transit geometries can also be eliminated, as the impact parameter is constrained to $b<0.7$ at the 99\% level based on our transit and contamination analyses. 
The apparent boxy shape of our follow-up lightcurves is in stark contrast with the V-shaped transit expected for grazing orbits. Hence, grazing EB scenario is ruled out.

Moreover, we can constrain the classes of HEBs that can reproduce the observed transit depth and shape using our multi-band observations. We aim to compute the eclipse depths for a range of plausible HEBs in the bluest and reddest bandpasses where they are expected to vary significantly.
We adopt the method presented in \citet{2020BoumaTOI837} to perform the calculation taking into account non-zero impact parameter, the details of which are discussed in the Appendix~\ref{sec:app-heb}.
Comparing the simulated eclipse depths with the observed depth in each band, we found that there is no plausible HEB configuration explored in our simulation that can reproduce the observed depths in multiple bands simultaneously. Hence, the HEB scenario is ruled out.

Although, \target's probability of being a BEB is very high a priori given its location at the galactic plane, we argue in the following that the BEB scenario is extremely unlikely.

Our MuSCAT3 observation can resolve the signal down to 3\arcsec, which represents the maximum radius within which the signal must originate. Furthermore, our high-resolution speckle imaging ruled any nearby star and blended sources down to 0.1\arcsec at a delta mag of 4.5. 
We checked archival images taken more than 60 years apart, but the proper motion of \target is not enough to obtain a clear view along the line of sight of the star. However, we can use statistical arguments to estimate the probability of a chance-aligned star. To do this, we use the population synthesis code \trilegal\footnote{\url{http://stev.oapd.inaf.it/cgi-bin/trilegal}} \citep{2005GirardiTrilegal}, which can simulate the Galactic stellar population along any line of sight. Given the position of \target, we found a probability of $5\times 10^{-8}$ to find a star brighter than $T$=16\footnote{$T$ denotes the \tess bandpass. The maximum delta magnitude was computed using dT=-2.5$\log_{10}$(depth), which translates to the magnitude that can produce a 100\% eclipse}, within an area equal to the smallest MuSCAT3 photometric aperture (aperture radius $=$ 3\arcsec).
Assuming all such stars are binary and preferentially oriented edge-on to produce eclipses with period and depth consistent with the \tess detection, then this can represent a very conservative upper limit of a BEB scenario. Despite the small probability of a BEB based on the trivial star counting argument, we discuss relevant tools in the following section for a more thorough statistical modeling.

\subsection{Statistical validation} \label{sec:statistical}
Here we quantify the false positive probability (FPP) of \target.01 using the Python package \vespa and \triceratops 
\citep{2015MortonVespa, 2020GiacaloneDressing}, the details of which are discussed in Section~\ref{app:validation}.
Although we were able to rule out the classes of EB, BEB, and HEB in Section~\ref{sec:validation}, we ran \vespa considering all these scenarios for completeness and computed a formal FPP$<1\times10^{-6}$ 
which robustly quantifies \target.01 as a statistically validated planet. 
Additionally, we validated \target.01 using \triceratops and found FPP=0.0020. \citet{2021GiacaloneTOI} noted that TOIs with FPP~$ < 0.015$ have a high enough probability of being bona fide planets to be considered validated. The low FPPs calculated using \vespa and \triceratops added further evidence to the planetary nature of \target.01. We now refer to the planet as \target\,b in the remaining sections.

\section{Discussion} \label{sec:discussion}

\begin{figure}
    \centering
    \includegraphics[width=0.48\textwidth]{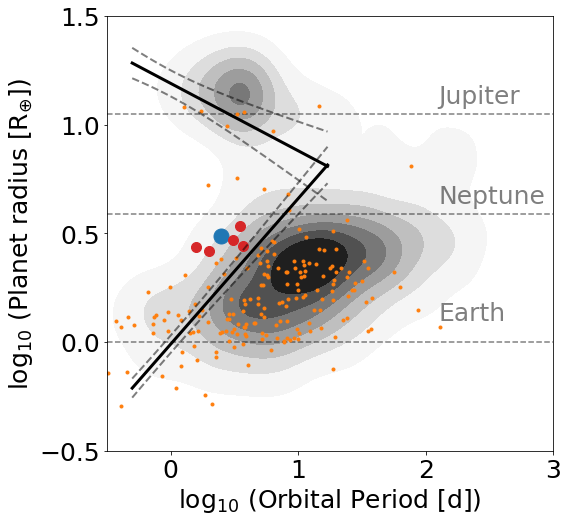}
    \caption{\target\,b (blue) in the context of known transiting planets (contours). \target\,b appears to be within or close to the boundaries of the Neptunian desert (solid black lines) in the period-radius plane defined by \citet{Mazeh2016}. The dashed lines refer to the boundaries' uncertainty regions. The orange points indicate planets orbiting M dwarfs (\teff$<3800$K) and the red points indicate the five planets most similar in this parameter space to \target\,b: K2-25\,b, K2-320\,b, GJ 1214\,b, TOI-269\,b, and TOI-2406\,b.}
    \label{fig:rad-per}
\end{figure}

\subsection{The nature of the planet}

Here, we consider the nature of \target\,b by placing it in context with the population of known exoplanets\footnote{Based on a query of the NASA Exoplanet Archive ``Confirmed Planets'' table on 2022 January 31, \url{https://exoplanetarchive.ipac.caltech.edu/}}.
Figure~\ref{fig:rad-per} shows a radius vs period diagram, indicating that there are only a handful of planets with similar characteristics to \target\,b. 
The measured planetary radius \rpl of \prad~\rearth and the orbital period $P$ of \pper~days, places it securely within the bounds of the Neptunian desert as defined by \citet{Mazeh2016}. The region occupied by \target remains sparsely populated despite recent discoveries of \tess planets within the Neptunian desert \citep[e.g.][]{2021Murgas, 2022Brande}.

It should be noted that the Neptunian desert was originally determined based on a population of planets orbiting mainly solar-type stars from the \kepler mission. 
Because \target is an M dwarf, the incident flux at a given orbital separation will be less than solar-type stars. 
Nevertheless, we emphasize that the target exists in a sparsely populated region of parameter space, despite the large number of planets discovered around M dwarfs since the \kepler mission (i.e. from \ktwo and \tess). For example, if we limit the comparison to the 279 confirmed planets around M dwarfs with \teff below 3800 K, only 14 planets have been found so far with orbital periods shorter than 10 days and planetary radii in the range 2.5\rearth~$<$\rpl$<5$~\rearth. 
As shown in Figure~\ref{fig:rad-per}, \target\,b is similar to K2-25 b\citep{Mann_K2-25}, K2-320 b\citep{Castro_K2-320}, GJ 1214 b\citep{Charbonneau_GJ1214}, TOI-269 b\citep{Cointepas_TOI-269}, and TOI-2406 b\citep{wells2021} in terms of orbital period and radius. In particular, TOI-2406\,b appears most similar to \target\,b as it orbits around a mid-M dwarf with an effective temperature of $3100 \pm 75 $, and has a radius of $2.94 \pm 0.17$~\rearth and orbital period of 3.077 days. TOI-2406 is also thought to be relatively old without any activity signal. As both \target\,b and TOI-2406\,b are excellent targets for detailed characterization studies, together they may provide unique insights into this class of planet. There is also some similarity between \target\,b and the Neptunian Desert planets orbiting young host stars, such as AU Mic\,b and\,c, K2-25\,b, K2-95\,b and K2-264\,b. It has been suggested that these planets may have inflated radii and could possibly still be undergoing atmospheric mass-loss \citep[e.g.][]{Mann_K2-25}. Further study of \target\,b could reveal whether its similarity to these planets (despite being older) is only superficial, or if it is indicative of an inflated radius.

\subsection{Prospects for transmission spectroscopy}

Given the rarity of this planet, it would be useful to assess its prospects for future atmospheric observations to understand its formation and evolution. In particular, the relatively large size of the planet compared to its host star makes it a good candidate for transmission spectroscopy.
Using Equation~1 in \citet{Kempton2018}, we calculated the transmission spectroscopy metric (TSM) of \target\,b from its mass, radius, equilibrium temperature, stellar radius, and $J$-band magnitude. We used the values in Figure~\ref{tab:stellar} and \ref{tab:results}, and assumed a mass of \pmass\,\mearth estimated by \mrexo. The derived TSM value of \target\,b is \ptsm. For reference, \citet{Kempton2018} suggested that planets with TSM$>$90 are ideal targets for atmospheric follow-up.

\begin{figure*}
    \centering
    \includegraphics[width=0.67\textwidth]{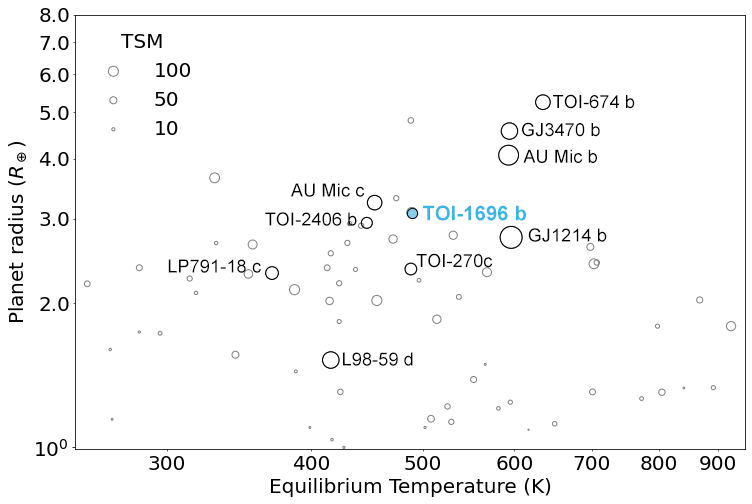}
    \caption{Planetary equilibrium temperature vs radius for \target\,b and other planets with $Rp < 10R_\oplus$, with the host stars having \teff$<3800\rm{K}$ and $H<11\rm{mag}$. The point size represents the calculated TSM values. Data points with a planet name beside them are those with a higher TSM values than the target.}
    \label{fig:tsm}
\end{figure*}

For comparison, we calculated the TSM for the known population of transiting M dwarf planets. We selected planets with \teff~$<3800\rm{K}$, \rpl $< 10R_\oplus$, and $\rm{H}<11\rm{mag}$\footnote{Based on a query of the NASA Exoplanet Archive Confirmed Planets table as of 2022 January 31}. For planets without mass measurements, we assumed the masses predicted by \mrexo. For planets without an equilibrium temperature, we estimated it from the semi-major axis and the host star's effective temperature (assuming zero albedo). Figure \ref{fig:tsm} shows the computed TSM values for the selected samples of planets. The TSM of \target\,b places it in the top 10, making it one of the best targets for future atmospheric investigations.

\begin{figure}
    \centering
    \includegraphics[width=\columnwidth]{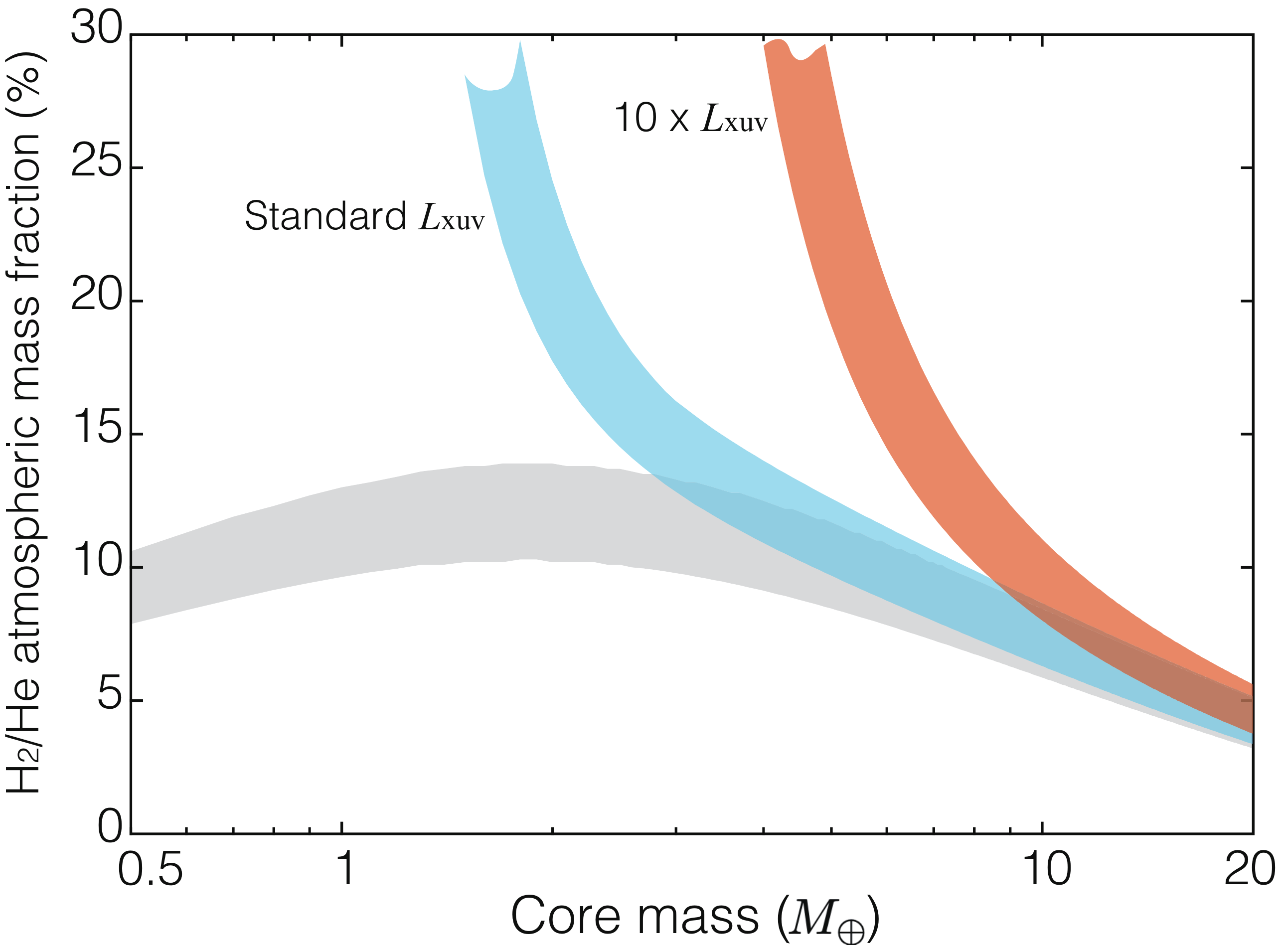}
    \caption{Initial H$_2$/He atmospheric mass fraction of a \target\,b-like planet that satisfies the radius of \prad $R_\oplus$ and $T_\mathrm{eq} = 489\pm13$\,K after photo-evaporative mass loss for $8$\,Gyr under the standard XUV radiation field ($L_\mathrm{XUV}$) and $10 L_\mathrm{XUV}$. The grey region shows the H$_2$/He atmospheric mass fraction that reproduces the observed radius of \target\,b with a rocky core.}
    \label{fig:McMHHe}
\end{figure}

\subsection{Existence of a primordial atmosphere}

Up to this point in the section, the discussion has been based on the assumption that the target has an atmosphere. Usually it is thought that planets above the so-called radius gap can retain their atmospheres \citep{2014ApJ...783L...6W,2015ApJ...801...41R}. However, does \target\,b actually have an H$_2$/He atmosphere? Here we study the atmospheric mass that \target\,b can retain after $\sim 8$\,Gyr under a stellar XUV irradiation. The mass of \target\,b remains poorly constrained as discussed in Section~\ref{sec:mass}. We modeled \target\,b as a rocky planet with Earth-like core compositions (MgSiO$_3$:Fe = 7:3) in the mass range from $0.5M_\oplus$ to $20M_\oplus$. The silicate mantle and iron core were described by the 3rd-order Birch-Murnagham EoS for MgSiO$_3$ perovskite \citep{2000PhRvB..6214750K,2007ApJ...669.1279S} and the Vinet EoS for $\epsilon$-Fe \citep{2001GeoRL..28..399A}, respectively. The Thomas-Fermi Dirac EoS \citep{1967PhRv..158..876S} was applied to high-pressure EoS for MgSiO$_3$ at $P \geq 4.90$\,TPa and Fe at $P \geq 2.09 \times 10^4$\,GPa \citep{2007ApJ...669.1279S,2013PASP..125..227Z}. The pressure and temperature in a H$_2$/He envelope were calculated using the SCvH EoS \citep{1995ApJS...99..713S}.

We computed the thermal evolution of \target\,b with a H$_2$/He atmosphere by calculating its interior structure in hydrostatic equilibrium for $\sim 8$\,Gyr, and calculated its mass loss process. 
The initial mass fraction of a H$_2$/He atmosphere for a rocky planet ranges from 0.001\% to 30\% of its core mass.
The energy-limited hydrodynamic escape \citep{1981Icar...48..150W} controls the mass loss rate given by 
\begin{equation}
  \frac{dM_\mathrm{p}}{dt} = - \eta \frac{R^3_\mathrm{p} L_\mathrm{XUV}(t)}{4  G  M_\mathrm{p}  a^2  K_\mathrm{tide}},
  \label{eq:Mdot}
\end{equation}
$\eta$ is the heating efficiency due to stellar XUV irradiation, $L_\mathrm{XUV}$ is the stellar XUV luminosity, $G$ is the gravitational constant, and $R_\mathrm{p}$ is the planetary radius \citep{2007A&A...472..329E}. Since the heating efficiency for a hydrogen-rich upper atmosphere was lower than 20\,\% \citep{2014A&A...571A..94S,2015SoSyR..49..339I}, we adopted $\eta = 0.1$. 
$K_\mathrm{tide}$ is the reduction factor of a gravitational potential owing to the effect of a stellar tide:
\begin{equation}
  K_\mathrm{tide}(\xi) = 1 - \frac{3}{2\xi} + \frac{1}{2\xi^3} < 1,\,\,\,\,\,\xi = \frac{R_\mathrm{H}}{R_\mathrm{p}},
  \label{eq:Ktide}
\end{equation}
where $R_\mathrm{H}$ is the Hill radius.
The XUV luminosity ($L_\mathrm{XUV}$) of \target followed  from the X-ray-to-bolometric luminosity relations of M-type stars \citep{2012MNRAS.422.2024J}, where we adopted the current luminosity of \target as its bolometric luminosity.
We also considered a $10 L_\mathrm{XUV}$ model because of the large uncertainty in $L_\mathrm{XUV}$ of young M dwarfs.

Figure \ref{fig:McMHHe} shows the initial H$_2$/He atmosphere of a \target\,b-like planet that reproduces the radius of \prad $R_\oplus$ at the current location (i.e., $T_\mathrm{eq} = 489\pm13$\,K) after the mass loss driven by the standard XUV radiation ($L_\mathrm{XUV}$: blue) and 10 times higher one ($10 L_\mathrm{XUV}$: red).
The grey region shows the H$_2$/He atmospheric mass fraction of \target\,b with a rocky core that satisfies its observed radius.
The observed radius of \target\,b favors the existence of a H$_2$/He atmosphere atmosphere with $\gtrsim 3$\,wt\% unless its core contains icy material. We find that \target\,b can possess the H$_2$/He atmosphere for 8\,Gyr if its core mass is larger than $\sim 1.5M_\oplus$ ($\sim 4M_\oplus$ for 10$L_\mathrm{XUV}$ models).
If \target\,b initially had the H$_2$/He atmosphere of $\lesssim 3\%$, it should be completely lost. Also, \target\,b with mass of $\gtrsim 10M_\oplus$ can retain almost all the H$_2$/He atmosphere accreted from a disk.
These suggest that \target\,b with a rocky core of $\gtrsim 1.5-4M_\oplus$ is likely to be a sub-Neptune with a H$_2$/He atmosphere.

\section{Conclusions} 
\label{sec:conclusion}

\tess found transit signals of a sub-Neptune planet orbiting a mid-M dwarf \target. To validate and characterize the planetary system, we conducted follow-up observations of this system including ground-based transit photometry, high-resolution imaging, and high- and medium-resolution spectroscopy. 

We have used several methods to determine the stellar parameters based on the results of the spectroscopic observations, and have confirmed that the results are consistent. The host star, \target, is a M-type star with a \mstar at \smass~\msun and \teff at \steff~K.

The fact that this target is located near the Galactic plane makes validation difficult. We used the results obtained to rule out various scenarios that could reproduce the \tess signal (grazing EB, HEB, and BEB). 

The validated planet, \target\,b is a Sub-Neptune size planet with the radius at \pradvalue~\rearth and rotation period at \ppervalue~days, which locates in the Neptunian desert. To see its atmospheric properties, we calculated how much of the atmosphere it currently retains, and found the planet likely to retain the H$_2$/He atmosphere if it has a core of $>$ 1.5--4$\rm{M_\oplus}$. In order to statistically evaluate the feasibility of transmission spectroscopy on this planet, we have also calculated and compared the TSM and concluded that this target is one of the planets with the best prospects for atmospheric detection among the currently known Sub-Neptune-sized planets. In addition, future RV observations with high-resolution infrared spectrographs such as IRD will allow us to place more substantial limits on the planetary mass.

\section{Acknowledgements}
Funding for the \tess mission is provided by NASA's Science Mission Directorate. 
We acknowledge the use of public \tess data from pipelines at the \tess Science Office and at the \tess Science Processing Operations Center. 
This research has made use of the Exoplanet Follow-up Observation Program website, which is operated by the California Institute of Technology, under contract with the National Aeronautics and Space Administration under the Exoplanet Exploration Program.
Resources supporting this work were provided by the NASA High-End Computing (HEC) Program through the NASA Advanced Supercomputing (NAS) Division at Ames Research Center for the production of the SPOC data products.
This paper includes data collected by the \tess mission that are publicly available from the Mikulski Archive for Space Telescopes (MAST).

This work makes use of observations from the Las Cumbres Observatory global telescope network. Some of the observations in the paper is based on observations made with the MuSCAT3 instrument, developed by Astrobiology Center and under financial supports by JSPS KAKENHI (JP18H05439) and JST PRESTO (JPMJPR1775), at Faulkes Telescope North on Maui, HI, operated by the Las Cumbres Observatory.

This research is in part on data collected at the Subaru Telescope, which is operated by the National Astronomical Observatory of Japan, and at the Gemini North telescope, located within the Maunakea Science Reserve and adjacent to the summit of Maunakea. We are honored and grateful for the opportunity of observing the Universe from Maunakea, which has cultural, historical, and natural significance in Hawaii.
Our data reductions benefited from PyRAF and PyFITS that are the products of the Space Telescope Science Institute, which is operated by AURA for NASA.
This research made use of Astropy,\footnote{http://www.astropy.org} a community-developed core Python package for Astronomy \citep{astropy:2013, astropy:2018}.
Some of the observations in the paper made use of the High-Resolution Imaging instrument(s) ‘Alopeke. ‘Alopeke was funded by the NASA Exoplanet Exploration Program and built at the NASA Ames Research Center by Steve B. Howell, Nic Scott, Elliott P. Horch, and Emmett Quigley. ‘Alopeke (and/or Zorro) was mounted on the Gemini North (and/or South) telescope of the international Gemini Observatory, a program of NSF’s NOIRLab, which is managed by the Association of Universities for Research in Astronomy (AURA) under a cooperative agreement with the National Science Foundation. on behalf of the Gemini partnership: the National Science Foundation (United States), National Research Council (Canada), Agencia Nacional de Investigación y Desarrollo (Chile), Ministerio de Ciencia, Tecnología e Innovación (Argentina), Ministério da Ciência, Tecnologia, Inovações e Comunicações (Brazil), and Korea Astronomy and Space Science Institute (Republic of Korea).
Some of the observations in this paper made use of the Infrared Telescope Facility, which is operated by the University of Hawaii under contract 80HQTR19D0030 with the National Aeronautics and Space Administration. The IRTF observations were collected under the program 2020B115 (PI: S. Giacalone).
This work has made use of data from the European Space Agency (ESA) mission
{\it Gaia} (\url{https://www.cosmos.esa.int/gaia}), processed by the {\it Gaia}
Data Processing and Analysis Consortium (DPAC,
\url{https://www.cosmos.esa.int/web/gaia/dpac/consortium}). Funding for the DPAC
has been provided by national institutions, in particular the institutions
participating in the {\it Gaia} Multilateral Agreement.
This work is supported by Grant-in-Aid for JSPS Fellows, Grant Number JP20J21872, JSPS KAKENHI Grant Numbers JP20K14518, JP19K14783, JP21H00035, JP18H05439, JP18H05439, JP20K14521, JP17H04574, JP21K20376, JP21K13975, JP18H05442, JP15H02063, and JP22000005, SATELLITE Research from Astrobiology Center (Grant Number AB031010, AB022006, and AB031014), and JST CREST Grant Number JPMJCR1761.

E. E-B. acknowledges financial support from the European Union and the State Agency of Investigation of the Spanish Ministry of Science and Innovation (MICINN) under the grant PRE2020-093107 of the Pre-Doc Program for the Training of Doctors (FPI-SO) through FSE funds.

\appendix
\renewcommand\thefigure{\thesection.\arabic{figure}}
\renewcommand\thetable{\thesection.\arabic{table}}

\section{Detailed methods of stellar parameter estimation}
\subsection{Abundances of eight metal elements from IRD spectra}
\label{sec-ap:ird_abundances}
\begin{figure*}
    \centering
    \includegraphics[width=0.8\textwidth]{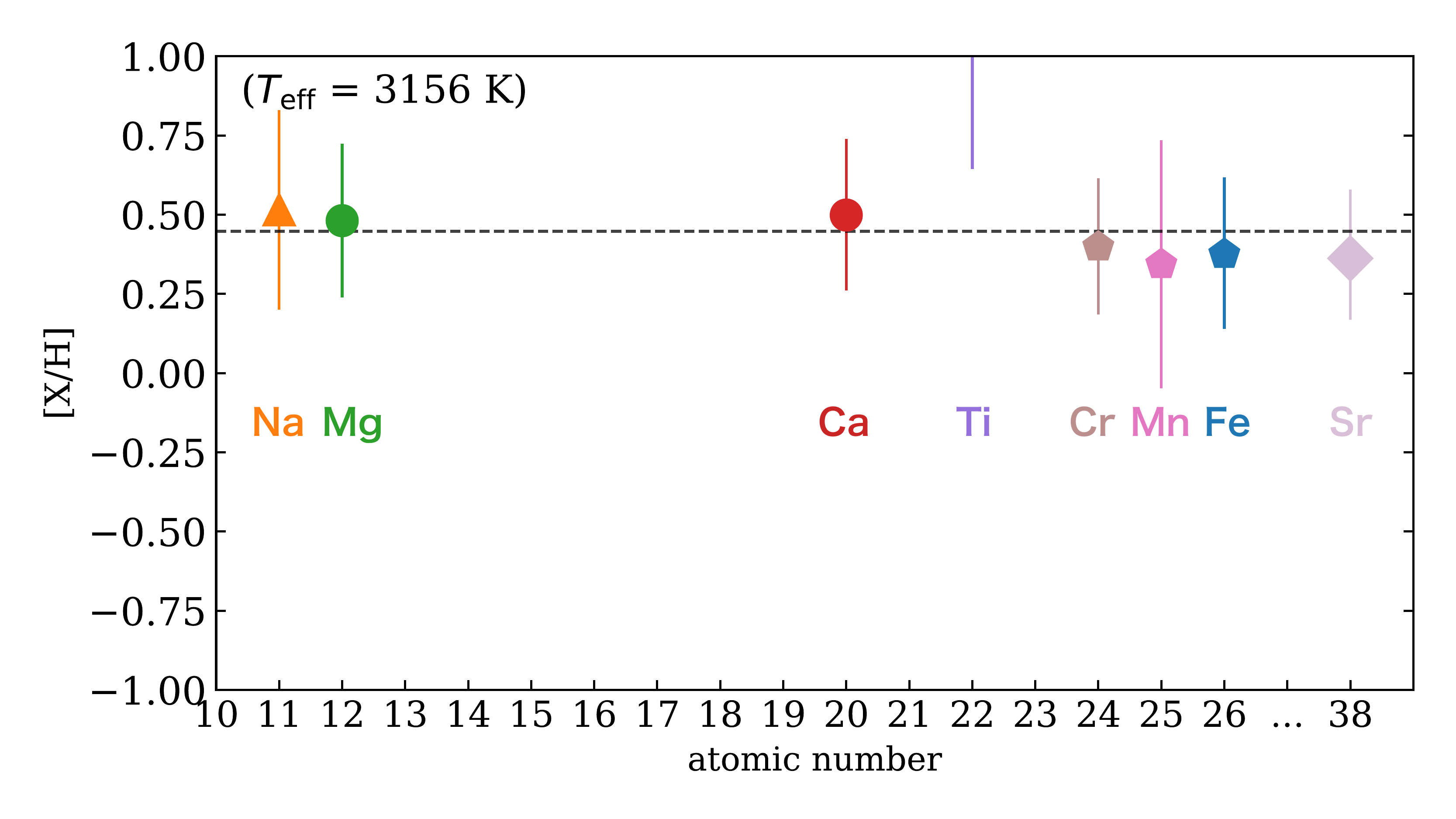}
    \caption{Metallicity values derived from IRD spectrum. The horizontal dashed line corresponds to the weighted average [M/H].}
    \label{fig:metallicity}
\end{figure*}

We calculated the abundances of seven other elements other than iron from IRD spectra. We used 28 lines in total caused by neutral atoms of Na, Mg, Ca, Ti, Cr, Mn, and Fe and singly ionized Sr.
The detailed procedures of abundance analysis and error estimation are described in \citet{Tako2020}.
Figure~\ref{fig:metallicity} shows the final values of abundance after the iteration. From the final values of the abundances of the eight elements, [M/H] was determined by calculating the average weighted by the inverse of the square of their estimated errors.

\subsection{Estimation of Stellar Radius and Mass: SED Fitting}
\label{sec-ap:stellarparam-SED}
\begin{figure}
    \centering
    \includegraphics[clip,trim={1cm 1cm 1cm 1cm},width=\columnwidth]{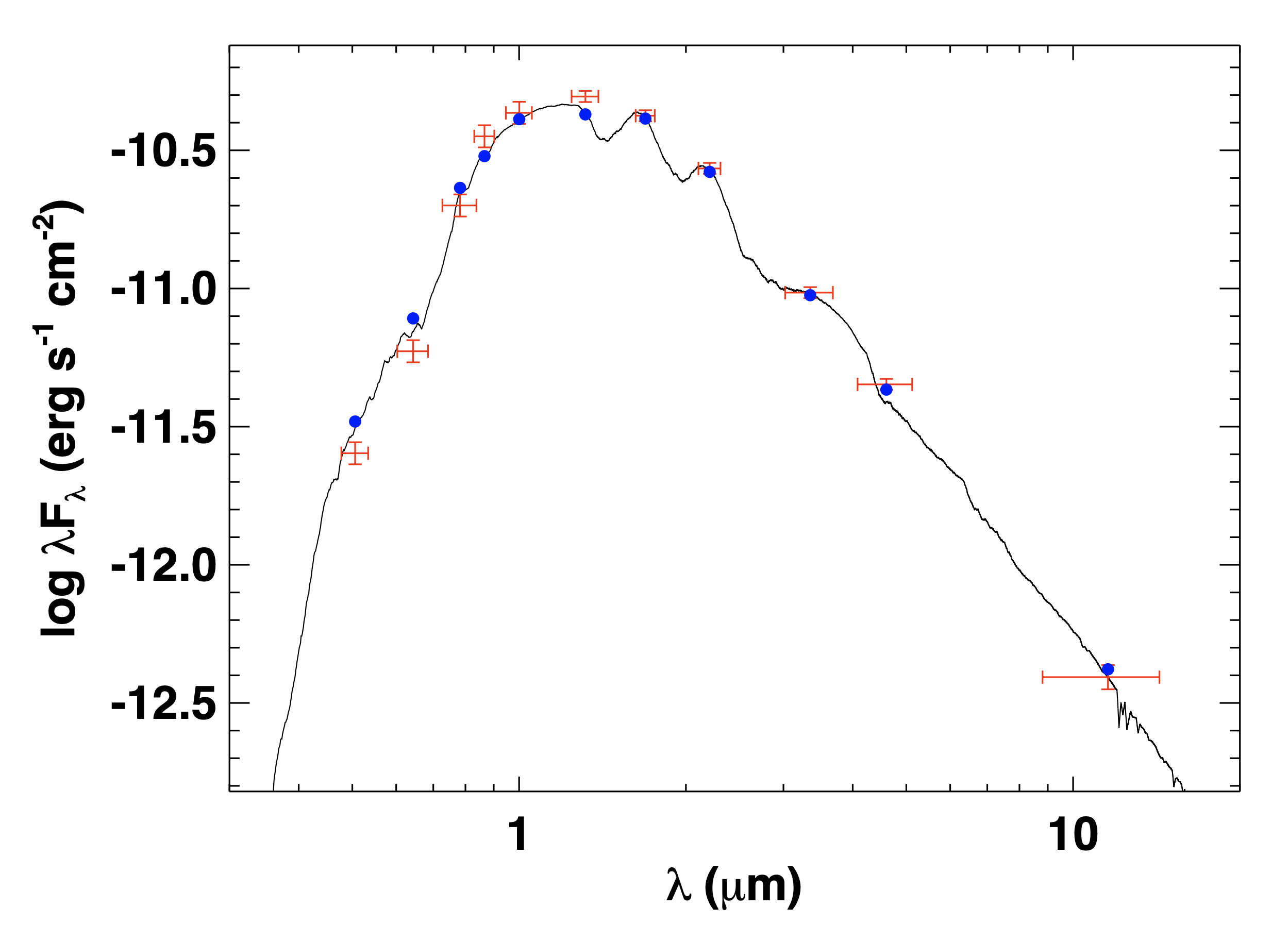}
    \caption{Spectral energy distribution of \target. Red symbols represent the observed photometric measurements, where the horizontal bars represent the effective width of the passband. Blue symbols are the model fluxes from the best-fit NextGen atmosphere model (black).  
    \label{fig:sed}}
\end{figure}

As an independent determination of the basic stellar parameters, we performed an analysis of the broadband spectral energy distribution (SED) of the star together with the\gaia EDR3 parallax \citep{StassunTorres:2021}, in order to determine an empirical measurement of the stellar radius, following the procedures described in \citet{Stassun:2016,Stassun:2017,Stassun:2018}. We pulled the $JHK_S$ magnitudes from {\it 2MASS}, the W1--W3 magnitudes from {\it WISE}, and the $grizy$ magnitudes from Pan-STARRS. Together, the available photometry spans the full stellar SED over the wavelength range 0.4--10~$\mu$m (see Figure~\ref{fig:sed}).  

We performed a fit using NExtGen stellar atmosphere models, with the effective temperature ($T_{\rm eff}$) and metallicity ([Fe/H]) constrained from the spectroscopic analysis. The remaining free parameter is the extinction $A_V$, which we fixed at zero due to the star's proximity. The resulting fit (Figure~\ref{fig:sed}) has a reduced $\chi^2$ of 1.7. Integrating the (unreddened) model SED gives the bolometric flux at Earth, $F_{\rm bol} = 5.20 \pm 0.25 \times 10^{-11}$ erg~s$^{-1}$~cm$^{-2}$. Taking the $F_{\rm bol}$ and $T_{\rm eff}$ together with the {\it Gaia\/} parallax, gives the stellar radius, $R_\star = 0.276 \pm 0.015$~R$_\odot$. We used the \teff and \feh values from spectroscopic results as priors for the parameter estimation.

In addition, we estimated the stellar mass from the empirical relations of \citet{Mann2019}, giving $M_\star = 0.279 \pm 0.014$~M$_\odot$. Finally, the radius and mass together imply a mean stellar density of $\rho_\star = 18.79 \pm 3.26$~g~cm$^{-3}$. 

\subsection{Stellar parameter comparison}
\label{sec-ap:stellarparam-isochrones}

In addition to the methods described above, we used the Python package \isochrones, which calculates stellar parameters from the stellar evolution models. The three methods are not fully independent, as some of them use the same relations such as mass derivation from \citet{Mann2019}, but comparing three results are useful to confirm the results are robust. The derived stellar parameters agreed within $1 \sim 2 \sigma$, as shown in \ref{tab:ap-stellar-comp}. We pick up the results from the empirical relations as our final stellar parameters in Table~\ref{tab:stellar}.
\setcounter{figure}{0} 
\setcounter{table}{0} 

\begin{table*}
\centering
\caption{Stellar parameters which were derived from empirical relations (Method 1; Section~\ref{sec:stellarparam-empirical}), SED fitting (Method 2; Section~\ref{sec-ap:stellarparam-SED}) and \isochrones (Method 3;Section~\ref{sec-ap:stellarparam-isochrones}). $^\dagger$The resulting posterior is approximately zero and non-Gaussian as a result of using a tight uniform prior close to 0 for numerical reasons.}
\label{tab:ap-stellar-comp}
\begin{tabular}{lccc}
Parameter & Method 1 & Method 2 & Method 3 \\
\hline
\teff (K) & - & $3130 \pm 75$ & $3159 \pm 40$  \\
\feh (dex) & - &  $0.2 \pm 0.3$ & $0.232^{+0.035}_{-0.038}$\\
\mstar (\msun) & \smass & $0.279 \pm 0.014$ & $0.276^{+0.006}_{-0.005}$ \\
\rstar (\rsun) & \srad & $0.280 \pm 0.014$  & $0.291 \pm 0.005$\\
\logg (cgs) & \slogg & $4.990 \pm 0.049$ & $4.955^{+0.008}_{-0.007}$ \\
\rhostar (\gcc) & \srho & $18.0 \pm 2.9$ & $16.282^{+0.598}_{-0.617}$ \\
distance (pc) & \sdist & - & $65.390^{+0.540}_{-0.464}$ \\
Luminosity (\lsun) & \slumi & - & - \\
$\mathrm{A_V}$ (mag) & 0 (fixed) &  0 (fixed) & $\sim$0$^\dagger$ \\
$F_{bol}$ (cgs) &- & $5.19(18) \times 10^{-11}$  & - \\
$M_Ks$ (mag) & $7.265 \pm 0.026$ & - & - \\
\hline
\end{tabular}
\end{table*}

\section{Contamination
Analysis}\label{app-contamination}
Contamination leads to a decrease in the observed transit depth (the planet appears to be smaller than it truly is), and this effect is achromatic even if the host and the contaminant(s) are of different spectral types. Having simultaneous multicolor photometry allows us to measure possible contamination and consequently provides strong constraints on the false positive scenarios discussed in Section~\ref{sec:validation}. 

Following the methods presented in \citet{2020ParviainenTOI263, 2021ParviainenTOI519}, we used the physics-based contamination model included in PyTransit v21 to model the light curves using a transit model that includes a light contamination component based on model stellar spectra leveraging multicolor photometry. Fitting the transit+contamination model to MuSCAT3 lightcurves allows us to measure the contamination in $i$-band\footnote{We adopt $i$ as reference passband for simplicity}, the effective temperature of the host (\teffh), and the effective temperature of the contaminant (\teffc).

We used normal priors for the period and $\rm{T_0}$ based on the results of our transit analysis. We also used normal priors on limb darkening, host effective temperature, and host star density, based on our spectroscopic analysis. Among them, the spectroscopic priors are the most important. Without a limb darkening prior, the transit fit in $g$-band is boxy perhaps due to the sparse data sampling. Without the \teffh prior, the posteriors are not well behaved. Without the host \rhostar prior, the model converges to very high values ($\sim$33\gcc) which is inconsistent with the results from our previous analyses. 

\begin{figure*}
    \centering
    \includegraphics[width=\textwidth]{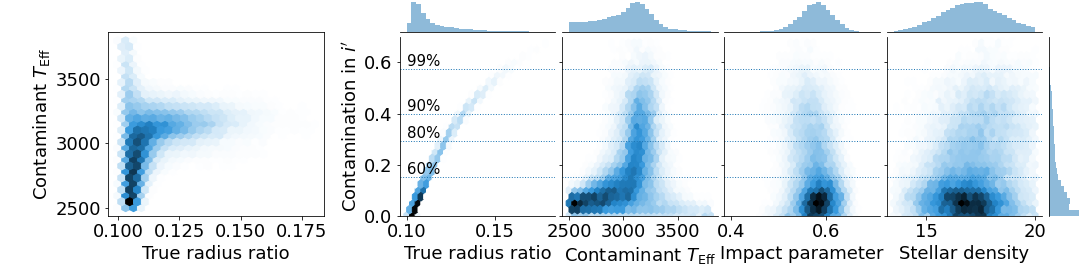}
    \caption{Joint and marginal posteriors for the key parameters from the transit+contamination modelling of the MuSCAT3 multicolor light curves. Contamination due to sources with significantly different effective temperature than the host is ruled out.}
    \label{fig:contamination}
\end{figure*}

The joint and marginal posteriors of the relevant parameters are shown in Figure~\ref{fig:contamination}.
Significant levels of blending from sources with effective temperature different from that of the host star are excluded, and also the blending from sources with \teffc $\sim$ \teffh are strongly constrained. 

\section{HEB simulation} \label{sec:app-heb}
We assumed that each system was composed of the primary star (\target, Star 1), plus a tertiary companion (Star 3) eclipsing a secondary companion (Star 2) every \ppervalue d. For a grid of secondary and tertiary star masses ranging from 0.1 to 0.4\msun, we then calculated the observed maximum eclipse depth caused by Star 3 eclipsing Star 2 in MuSCAT3 $g$- and $z$-bands using the following procedure. First, we interpolated \lstar and \teff of Star 2 and Star 3 from MIST isochrones given their masses, and the age, metallicity, and mass of Star 1 in Table~\ref{tab:stellar}. We then computed the blackbody function of each star given their \teff then convolved it with the transmission functions for each band downloaded from the SVO filter profile service\footnote{\url{http://svo2.cab.inta-csic.es/theory/fps/}}. We then integrated the result using the trapezoidal method and computed the bolometric flux \fbol, using the integrated functions above. Using 
Stefan-Boltzmann law and given \teff and \lstar, we computed the component radii and luminosities to derive the eclipse depth. 

\begin{figure*}
    \centering
    \includegraphics[clip,trim={0 0 0 0},width=\textwidth]{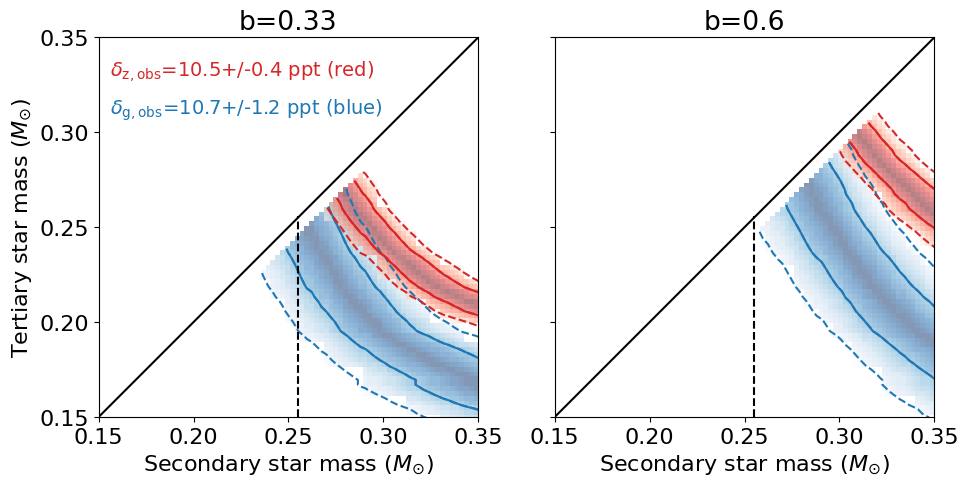}
    \caption{HEB mass configurations which produce eclipse depths in $g$-band (blue) and $z$-band (red) consistent with the observed depths (indicated in the upper left corner of the first panel). The left panel corresponds to the lower limit of the impact parameter and the right for the median value. The colored solid line and dashed lines correspond to confidence regions that are consistent with the observed depths within 1- and 2-$\sigma$, respectively. The vertical black line corresponds to the mass of the central star (i.e. \target). The fact that the red and blue regions do not overlap within 1-$\sigma$ taking into account impact parameter rules out the HEB false positive scenario. 
    }
    \label{fig:heb_joint}
\end{figure*}

Figure~\ref{fig:heb_joint} shows the HEB configurations that produce eclipse depths in $g$- (blue) and $z$-bands (red) that are consistent with the observed depth for two given impact parameters. 
The lower impact parameter corresponds to the 3-$\sigma$ lower limit derived from our contamination analysis while the other impact parameter corresponds to the median value derived in our transit analysis. 
We confirm that indeed eclipses of an HEB are always deeper in the red than in the blue bands (i.e higher $m_2$/$m_1$ in $z$- than $g$-band) since the eclipsing companions are usually redder than the central star.
The important point here is that the HEB configurations that produce eclipses consistent with our observation do not overlap within 1-$\sigma$ in $g$- and $z$-bands for any reasonable impact parameters. Note also that our contamination analysis constrained possible contaminants to have the same colour as the host star, so only masses very close to \target (vertical dashed line in Figure~\ref{fig:heb_joint}) are allowed. Thus, we can rule out the HEB false positive scenario.

\section{Validation with \vespa and \triceratops} \label{app:validation}

\vespa\footnote{\url{https://github.com/timothydmorton/VESPA}} was originally developed as a tool for statistical validation of planet candidates identified by the \kepler mission \citep[e.g.][]{2016Morton}, but has also been used extensively to validate planets from subsequent missions, such as \ktwo \citep[e.g. ][]{2018Livingston60planets,2021deLeonK2}. \vespa compares the likelihood of a planetary scenario to the likelihoods of several astrophysical false positive scenarios involving eclipsing binaries (EBs), hierarchical triple systems (HEBs), background eclipsing binaries (BEBs), and the double-period cases of all these scenarios. The likelihoods and priors for each scenario are based on the shape of the transit signal, the star's location in the Galaxy, and single-, binary-, and triple-star model fits to the observed photometric and spectroscopic properties of the star generated using \isochrones. 
We used the MuSCAT3 lightcurve because of its high SNR and low levels of limb darkening, which provides the best constraint on the transit shape.
We also used the Gemini and Palomar contrast curves described in Section~\ref{sec:speckle}, a maximum aperture radius of \maxrad=3\arcsec (interior to which the transit signal must be produced), and ran the simulation using a population size of n=$10^6$, resulting to a formal FPP$<1\times10^{-6}$.

We also used \triceratops\footnote{\url{https://github.com/stevengiacalone/triceratops}} which is a tool developed to validate TOIs \citep{2020GiacaloneDressing, 2021GiacaloneTOI} by calculating the Bayesian probabilities of the observed transit originating from several scenarios involving the target star, nearby resolved stars, and hypothetical unresolved stars in the immediate vicinity of the target. These probabilities were then compared to calculate a false positive probability (FPP; the total probability of the transit originating from something other than a planet around the target star) and a nearby false positive probability (NFPP; the total probability of the transit originating from a nearby resolved star).
Given our follow-up photometry rules out nearby stars as a potential source of the transit signal, we eliminate all stars except the target in the \triceratops analysis.
As an additional constraint, we use the contrast curve from our follow-up speckle imaging as a direct input in \triceratops. For the sake of reliability, we performed the calculation 20 times for the planet candidate and found FPP=0.0020. 
The low FPPs calculated using \vespa and \triceratops are small enough to statistically validate \target.01 as a planet.

\clearpage
\bibliography{ref}{}
\bibliographystyle{aasjournal}
\end{document}